\documentclass[aps,prb,twocolumn,superscriptaddress,
groupedaddress,10pt]{revtex4}
\usepackage{amsmath}
\usepackage{amssymb}
\usepackage{graphicx}
\usepackage{epsfig}
\usepackage{dcolumn}
\usepackage{bm}
\usepackage{bm}
\usepackage{blindtext}
\usepackage{natbib}
\usepackage{siunitx} 
\usepackage{multirow}

\usepackage[titletoc]{appendix}

\usepackage{simplewick}

\usepackage{bbm}
\makeatletter
\newcommand{\mathleft}{\@fleqntrue\@mathmargin0pt}
\newcommand{\mathcenter}{\@fleqnfalse}
\makeatother

\usepackage[usenames,dvipsnames]{xcolor}
\usepackage{amsthm}
\usepackage{booktabs}
\usepackage{hyperref}
\usepackage{tikz}
\usetikzlibrary{calc}
\usepackage[printwatermark]{xwatermark}
\usepackage{mathtools}

\def\be{\begin{equation}} \def\ee{\end{equation}}
\def\bea{\begin{eqnarray}} \def\eea{\end{eqnarray}}

\def\nn{\nonumber}

\begin{document}
\title{
Electric field effects in one-dimensional spin-1/2 $K_1J_1\Gamma_1\Gamma_1^\prime K_2J_2$ model with ferromagnetic Kitaev coupling
}

\author{Wang Yang}
\affiliation{School of Physics, Nankai University, Tianjin 300071, China}

\author{Helin Wang}
\affiliation{School of Physics, Nankai University, Tianjin 300071, China}

\author{Chao Xu}
\affiliation{Institute for Advanced Study, Tsinghua University, Beijing 100084, China}
\affiliation{Kavli Institute for Theoretical Sciences, University of Chinese Academy of Sciences, Beijing 100190, China}

\begin{abstract}

We perform a systematic study on the effects of electric fields in the Luttinger liquid phase of the one-dimensional spin-$1/2$ $K_1J_1\Gamma_1\Gamma_1^\prime K_2J_2$ model in the region of ferromagnetic nearest-neighboring Kitaev coupling. 
We find that while electric fields along $(1,1,1)$-direction maintain the Luttinger liquid behavior,
fields along other directions drive the system to a dimerized state.
An estimation is made on how effective a $(1,1,1)$-field is for tuning the Luttinger parameter in real materials. 
Our work is useful for understanding the effects of electric fields in one-dimensional generalized Kitaev spin models,
and provides a starting point for exploring the electric-field-related physics in two dimensions based on a quasi-one-dimensional approach.

\end{abstract}
%\pacs{75.10.Pq, 05.30.Rt, 71.10.Hf, 75.10.Jm}
\maketitle
%\tableofcontents

%%%%%%%%%%%%%%%%%%%%%%%%%%%%%%%%%%%%%%%
\section{Introduction}

Kitaev materials are two-dimensional (2D) solid state systems \cite{Jackeli2009,Chaloupka2010,Singh2010,Price2012,Singh2012,Plumb2014,Kim2015,Winter2016,Baek2017,Leahy2017,Sears2017,Wolter2017,Zheng2017,Rousochatzakis2017,Kasahara2018,Rau2014,Ran2017,Wang2017,Catuneanu2018,Gohlke2018,Liu2011,Chaloupka2013,Johnson2015,Motome2020} that are considered to be potentially useful for realizing Kitaev spin-1/2 model \cite{Kitaev2006} on the honeycomb lattice, which is a prototypical exactly solvable model for realizing topological quantum computation \cite{Kitaev2006,Nayak2008}.  
Iridates and $\alpha$-RuCl$_3$ are typical Kitaev materials on the honeycomb lattice,
which both have a ferromagnetic (FM) Kitaev coupling. 
Although the exactly solvable Kitaev model is a spin liquid model without any magnetic order, existing Kitaev materials are all magnetically ordered at sufficiently low temperatures \cite{Hermanns2018,Takagi2019,Trebst2022,Matsuda2025}, including zigzag, counter-rotating spiral and triple-Q orders, which hinder their applications for the purpose of topological quantum computing.

A central theme in the field of Kitaev materials is to find a method that can tune the real materials into the Kitaev spin liquid phase. 
Electric and magnetic controls are common and convenient controls of  2D materials.
For example, experiments have found evidence that certain Kitaev materials exhibit half-quantized Hall conductance when external magnetic fields are applied above a critical value \cite{Kasahara2018,Yokoi2021Science,Bruin2022NatPhys,Czajka2021NatPhys},
which is a hallmark of the Kitaev spin liquid phase. 
On the other hand, electric field tuning of Kitaev materials  remains much less explored in experiments. 

On the theory side, the deviations of Kitaev materials away from the idealized spin liquid behaviors have been identified to originate from the existence of non-Kitaev coupling in real materials, 
including Heisenberg coupling \cite{Chaloupka2010,Jackeli2009}, the off-diagonal Gamma and $\Gamma^\prime$ couplings \cite{Plumb2014}, and beyond nearest-neighboring  interactions.  
Models that include these additional couplings are termed as generalized Kitaev models.
For example, theories and experiments have revealed that the couplings satisfy $(K_1<0,J_1>0,\Gamma_1>0)$ in honeycomb  iridates
and $(K_1<0,J_1<0,\Gamma_1>0)$ in $\alpha$-RuCl$_3$ \cite{Liu2022},
where $K_1$, $J_1$ and $\Gamma_1$ denote the nearest neighboring Kitaev, Gamma, and Heisenberg interactions, respectively. 
However, theoretical and numerical studies of generalized Kitaev models are very demanding because of their strongly correlated nature. 

In view of the aforementioned difficulties in 2D, a productive route is to study decoupled chains as a first step, leaving inter-chain interactions for a subsequent  perturbative treatment. 
The 1D systems have the particular advantage that they are usually amenable for both analytical and numerical studies,
because of various powerful  1D methods \cite{Haldane1981,Haldane1981a,Belavin1984,Knizhnik1984,Affleck1985,Affleck1988,Affleck1995a,White1992,White1993,Schollwock2011} including bosonization, conformal field theory, and density matrix renormalization group (DMRG) numerics. 
Along this line, there have been surging research interests in 1D generalized Kitaev spin models \cite{Sela2014,Agrapidis2018,Agrapidis2019,Catuneanu2019,Yang2020,Yang2020a,Yang2020b,Yang2021b,Yang2022,Yang2022b,Yang2020c,Yang2025a,Yang2025b,Yang2024,Luo2021,Luo2021b,You2020,Sorensen2021},
particularly focusing on effects of the bond-dependent Kitaev coupling and the off-diagonal Gamma coupling \cite{You2020,Liu2021IsingGammaPhysicaA,Zhao2022XYGammaPRA,Kheiri2024XYGammaPRB,Mahdavifar2024XXZUniformGammaSciRep,Abbasi2025ModulatedGammaSciPost,Jin2025XXZStaggeredGammaPRE,Abbasi2025XXGammaFieldsSciRep,Abbasi2025ConcurrenceRinP,SaitoHotta2024ZigzagGamma_arXiv,Haddadi2024XYGammaBattery}. 
Rich strongly correlated physics have been revealed in 1D Kitaev and Gamma models,
including nonsymmorphic symmetries \cite{Yang2024}, emergent SU(2)$_1$ conformal invariance \cite{Yang2020,Yang2025a}, exotic symmetry breaking phases, 
chiral soliton states \cite{Sorensen2023ChiralSoliton,Sorensen2023IntegerSpinSolitons}, topological domain-wall excitations \cite{LaurellAlvarezDagotto2023PRB_CompassDW}, and Lifshitz-type quantum phase transitions \cite{You2020,SaitoHotta2024ZigzagGamma_arXiv,Jin2025XXZStaggeredGammaPRE}.
It is also worth mentioning that besides providing hints for 2D physics,
1D studies have their independent merits as well,
since experiments have established the existence of real 1D Kitaev materials \cite{Morris2021TwistedKitaevNatPhys,Churchill2024}. 

Regarding electromagnetic  controls in 1D, existing theoretical works in generalized Kitaev spin chains and ladders have been mostly focusing on magnetic fields
\cite{Sorensen2023ChiralSoliton,Subrahmanyam2013KitaevTypeTF,Bhullar2025FieldInducedOrdersPRB,Birnkammer2024RamanConfinementPRB,MetavitsiadisBrenig2021FluxMobilityPRB,You2018CompassAltFieldPRB,Wu2019CompassZeroModesPRB},
with electric fields much less explored. 
In Ref. \onlinecite{Yang2025a}, the authors studied the effects of electric fields along the third direction in spin-1/2 Kitaev-Gamma chain in the limit of vanishing Hund's coupling,
where the first and second directions refer to the two bond directions in the decoupled 1D model, 
and the third direction is essentially the bond direction for inter-chain couplings within the parent 2D lattice. 
A systematic study of electric fields along general directions in more realistic  models is still lacking. 

In this work, we make a comprehensive study on the effects of static and uniform  electric fields in 1D spin-$1/2$ $K_1J_1\Gamma_1\Gamma^\prime_1K_2J_2$ model in the $(K_1<0,J_1>0,\Gamma_1>0)$ region, 
where the subscripts ``$1$" and ``$2$" in $K_1J_1\Gamma_1\Gamma^\prime_1K_2J_2$ are used to denote nearest and second-nearest neighboring couplings, respectively. 
The chosen parameter range is relevant for honeycomb iridates \cite{Liu2022},
and the model under consideration is a realistic one which includes up to second-nearest-neighboring interactions. 
In the absence of electric fields, the low energy physics of the  1D system can be described by Luttinger liquid theory
when the $\Gamma^\prime_1,K_2,J_2$ interactions are small enough, 
which are indeed much smaller than the $K_1,J_1,\Gamma_1$ couplings in real materials. 

For nonzero electric fields, we analyze the effects of the fields using a field theory perturbation based on the SU(2)$_1$ Wess-Zumino-Witten (WZW) model. 
We find that the $(1,1,1)$-direction is special: The system always remains in the Luttinger liquid phase when an electric field along 
the $(1,1,1)$-direction is applied, as long as the field is not so large that it goes beyond the perturbative region.   
An estimation is made  on how effective it is to use a $(1,1,1)$-electric-field for tuning  the Luttinger parameter in real materials.  
For all other directions of the electric fields, the system in general is driven into a dimerized state. 
Exceptions are electric fields along $(0,0,1)$- and $(1,1,0)$-directions in the limit of vanishing Hund's coupling,
under which circumstances the system retains its Luttinger liquid behavior. 
Analytical predictions are all consistent with our large-scale DMRG numerical simulations.

The rest of the paper is organized as follows. 
Sec. \ref{sec:model_Ham} gives the expressions for the Hamiltonian of the $K_1J_1\Gamma_1\Gamma^\prime_1K_2J_2$ model and the couplings induced by the electric fields.
Sec. \ref{sec:coupling_E_field} includes the expressions of the interactions in the Hamiltonian induced by electric fields. 
In Sec. \ref{sec:Low_energy_theory_zero_field}, the low energy field theory of the model in the absence of electric fields is briefly reviewed,
which provides a perturbative starting point for analyzing the effects of nonzero electric fields.
Sec. \ref{sec:E_fields_effects} discusses in details the effects of electric fields in various directions based on field theory perturbation,
which are confirmed numerically by DMRG simulations. 
Sec. \ref{sec:summary} summarizes the main results of the paper.

%%%%%%%%%%%%%%%%%%%%%%%%%%%%%%%%%%%%%%%%%%%%%%%%%%%%%
\section{Model Hamiltonian}
\label{sec:model_Ham}

In this section, we give the expression for the Hamiltonian of the spin-1/2 $K_1J_1\Gamma_1\Gamma^\prime_1K_2J_2$ chain. 
A useful unitary transformation called six-sublattice rotation is defined,
and symmetries of the model are discussed. 

%---------------------------------------------------------------------------------------------------------------------------- 
\subsection{The generalized Kitaev spin-1/2 chain}
\label{subsec:generalized_Kitaev}

%-------------------------------------------- 
\begin{figure}[ht]
\begin{center}
\includegraphics[width=8cm]{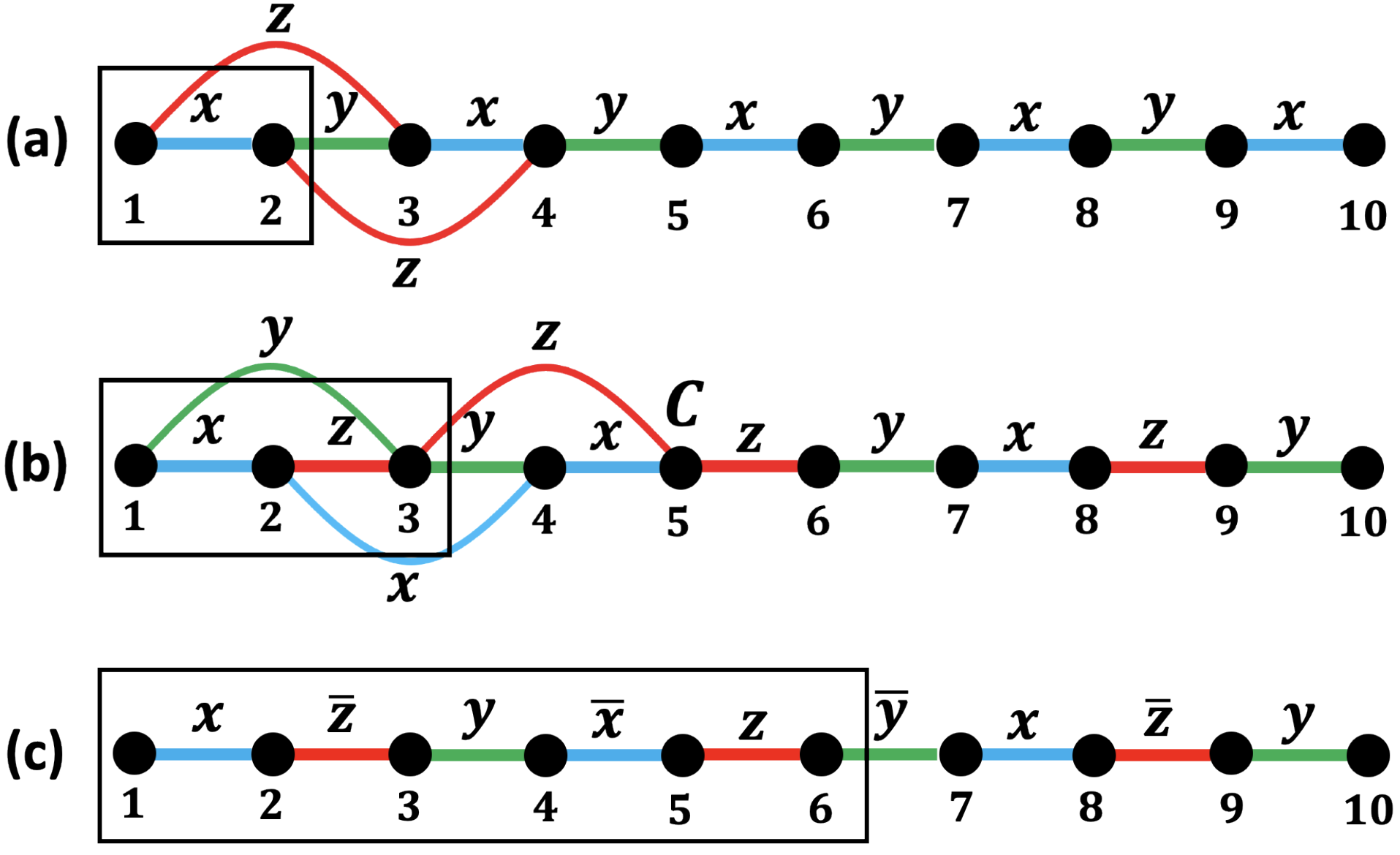}
\caption{Bond patterns of (a) the $K_1J_1\Gamma_1\Gamma^\prime_1K_2J_2$ chain without sublattice rotation,
(b) the $K_1J_1\Gamma_1\Gamma^\prime_1K_2J_2$ chain after the six-sublattice rotation,
(c) couplings induced by electric fields  after the six-sublattice rotation. 
Black squares represent unit cells for the bond patterns. 
} \label{fig:bond_pattern}
\end{center}
\end{figure}
%--------------------------------------------

The Hamiltonian of the spin-1/2 $K_1J_1\Gamma_1\Gamma^\prime_1K_2J_2$ chain is defined as
\begin{flalign}
&H=\sum_{\langle ij \rangle \in\gamma\,\text{bond}}\big[ K_1S_i^\gamma S_j^\gamma+ J_1\vec{S}_i\cdot \vec{S}_j+\Gamma_1 (S_i^\alpha S_j^\beta+S_i^\beta S_j^\alpha)\nn\\
&+\Gamma^\prime_1 (S_i^\gamma S_j^\alpha +S_i^\alpha S_j^\gamma+S_i^\gamma S_j^\beta+S_i^\beta S_j^\gamma)\big]\nn\\
&+\sum_{\langle \langle ij \rangle\rangle} \big[K_2 S_i^z S_j^z+J_2 \vec{S}_i\cdot \vec{S}_j\big],
\label{eq:Ham_1D}
\end{flalign}
in which 
$\langle ij \rangle$ and $\langle \langle ij\rangle\rangle$ are used to denote nearest and second nearest neighboring sites $i$ and $j$, respectively;
the bond pattern for $\gamma\in\{x,y\}$ is shown in Fig. \ref{fig:bond_pattern} (a),
$\alpha,\beta$ are the two spin directions among $\{x,y,z\}$ other than $\gamma$,
and $(\gamma,\alpha,\beta)$ forms a right-handed system. 
%In Eq. (\ref{eq:Ham_1D}), we have dropped the subscript ``1" in the coupling constants $K_1$, $J_1$, $\Gamma_1$, and $\Gamma^\prime_1$ for simplicity. 
Third-nearest neighboring terms are not included in the 1D model in Eq. (\ref{eq:Ham_1D}), 
since in 2D honeycomb lattice, there is only inter-chain but no intra-chain couplings in the leading third-nearest neighboring terms. 

In what follows, we parametrize $K_1,J_1,\Gamma_1$ as 
\bea
K_1&=&\sin(\theta)\cos(\phi)\nn\\
\Gamma_1&=&\sin(\theta)\sin(\phi)\nn\\
J_1&=&\cos(\theta). 
\label{eq:parametrize_KJG}
\eea
The condition $(K_1<0,\Gamma_1>0,J_1>0)$ is equivalent to $(0<\theta<\pi/2,\pi/2<\phi<\pi)$.
As discussed in Appendix \ref{app:estimations_E_couplings}, a  typical set of values for the couplings $K_1,J_1,\Gamma_1,\Gamma^\prime_1,J_2,K_2$ in iridates can be chosen as 
\bea
\theta&=&0.44\pi\nn\\
\phi&=&0.85\pi\nn\\
\Gamma_1^\prime&=&0.1\nn\\
K_2&=&0.02\nn\\
J_2&=&0.2,
\label{eq:numerics_KJG}
\eea
in which the normalization is chosen as $K_1^2+J_1^2+\Gamma_1^2=1$ in accordance with Eq. (\ref{eq:parametrize_KJG})
and the unit for the couplings is on the order of $10$ meV.
We will use the numbers in Eq. (\ref{eq:numerics_KJG}) for our DMRG numerical simulations in this work. 
We note that although Eq. (\ref{eq:numerics_KJG}) does not give the precise value of the couplings in real materials,
the relative orders of magnitude of the couplings are captured by the chosen numerical numbers Eq. (\ref{eq:numerics_KJG}).

It is straightforward to verify that the Hamiltonian in Eq. (\ref{eq:Ham_1D}) is invariant under the following symmetry transformations,
\begin{eqnarray}
1.&T &:  (S_i^x,S_i^y,S_i^z)\rightarrow (-S_{i}^x,-S_{i}^y,-S_{i}^z)\nn\\
%2.& T_{2a}&:  (S_i^x,S_i^y,S_i^z)\rightarrow (S_{i+2}^x,S_{i+2}^y,S_{i+2}^z)\nn\\
2.&T_a I_0&: (S_i^x,S_i^y,S_i^z)\rightarrow (S_{-i+1}^x,S_{-i+1}^y,S_{-i+1}^z)\nn\\
3.&R(\hat{n}_N,\pi)T_a&: (S_i^x,S_i^y,S_i^z)\rightarrow (-S_{i+1}^y,-S_{i+1}^x,-S_{i+1}^z),\nn\\
\label{eq:Sym_Neel_KHG}
\end{eqnarray}
in which $T$ is time reversal, $T_{na}$ ($n\in \mathbb{Z}$) is spatial translation by $n$ sites,
$I_0$ is spatial inversion with inversion center at site $0$,
$R(\hat{n},\theta)$ is global spin rotation around $\hat{n}$-axis with angle $\theta$,
and $\hat{n}_N=\frac{1}{\sqrt{2}}(1,-1,0)$.
The symmetry group $G$ of the system is generated by the symmetry operations in  Eq. (\ref{eq:Sym_Neel_KHG}) as
\begin{eqnarray}
G=\left< T,T_aI_0,  R(\hat{n}_N,\pi)T_a\right>.
\end{eqnarray}
Notice in particular that the two-site translation  $T_{2a}=[R(\hat{n}_N,\pi)T_a]^2$ is an element of $G$.

%---------------------------------------------------------------------------------------------------------------------------- 
\subsection{Six-sublattice rotation}
\label{subsec:six-rotation}

In real materials, the magnitudes of the coupling constants usually satisfy the following hierarchy
\bea
|K_2|,|J_2|,|\Gamma^\prime_1|\ll |J_1| \ll |K_1|,|\Gamma_1|.
\eea
Therefore, a good analytical strategy is to start with the nearest neighboring Kitaev-Gamma model, and then treat Heisenberg term $J_1$ as a perturbation, with small corrections from $\Gamma^\prime_1,K_2,J_2$ terms. 
However, the $\Gamma_1$ term has a rather complicated off-diagonal form,
involving interactions between spin operators along different directions. 
Therefore, a unitary transformation which maps the $\Gamma_1$ term to a diagonal form will be helpful for simplifying the analysis of the $\Gamma_1$ term. 
It turns out that such a unitary transformation indeed exists, acting as different spin rotations on different sites with a six-site periodicity,
which we discuss below.

The  six-sublattice rotation $U_6$ is defined as 
\begin{eqnarray}
\text{Sublattice $1$}: & (x,y,z) & \rightarrow (-x^{\prime},-y^{\prime},z^{\prime}),\nn\\ 
\text{Sublattice $2$}: & (x,y,z) & \rightarrow (x^{\prime},z^{\prime},-y^{\prime}),\nn\\
\text{Sublattice $3$}: & (x,y,z) & \rightarrow (-y^{\prime},-z^{\prime},x^{\prime}),\nn\\
\text{Sublattice $4$}: & (x,y,z) & \rightarrow (y^{\prime},x^{\prime},-z^{\prime}),\nn\\
\text{Sublattice $5$}: & (x,y,z) & \rightarrow (-z^{\prime},-x^{\prime},y^{\prime}),\nn\\
\text{Sublattice $6$}: & (x,y,z) & \rightarrow (z^{\prime},y^{\prime},-x^{\prime}),
\label{eq:6rotation}
\end{eqnarray}
in which "Sublattice $i$" ($1\leq i \leq 6$) represents all the sites $i+6n$ ($n\in \mathbb{Z}$) in the chain, and we have abbreviated $S^\alpha$ ($S^{\prime \alpha}$) as $\alpha$ ($\alpha^\prime$) for short ($\alpha=x,y,z$).
After the six-sublattice rotation, the Hamiltonian in Eq. (\ref{eq:Ham_1D}) becomes
\begin{flalign}
&H^\prime=\sum_{\langle ij \rangle\in\gamma\,\text{bond}}\big[-K_1S_i^{\prime\gamma} S_j^{\prime\gamma} + \Gamma_1 (S_i^{\prime\alpha} S_j^{\prime\alpha}+ S_i^{\prime\beta} S_j^{\prime\beta})\nn\\
&-J_1(S_i^{\prime\gamma} S_j^{\prime\gamma}+S_i^{\prime\alpha} S_j^{\prime\beta}+S_i^{\prime\beta} S_j^{\prime\alpha})\big]\nn\\
&-\sum_{\langle ij \rangle\in\gamma\,\text{bond}}\Gamma^\prime_1 \big[(S_i^{\prime\gamma} S_j^{\prime\alpha}-S_i^{\prime\alpha} S_j^{\prime\gamma})-(S_i^{\prime\gamma} S_j^{\prime\beta}-S_i^{\prime\beta} S_j^{\prime\gamma})\big] \nn\\
&+\sum_{\ \langle\langle ij\rangle\rangle\in\gamma_2\,\text{bond}}\big[
K_2S_i^{\prime\alpha_2}S_j^{\prime\beta_2}\notag\\
&+J_2(S_i^{\prime x}S_j^{\prime y}+S_i^{\prime y}S_j^{\prime z}+S_i^{\prime z}S_j^{\prime x})
\big],
\label{eq:Ham_rot}
\end{flalign}
in which the nearest-neighboring bond $\gamma$ and next-nearest-neighboring bond $\gamma_2$ have a three-site periodicity as shown in Fig. \ref{fig:bond_pattern} (b),
and both $(\gamma,\alpha,\beta)$ and $(\gamma_2,\alpha_2,\beta_2)$ form right-handed coordinate systems.

%---------------------------------------------------------------------------------------------------------------------------- 
\subsection{Symmetries}
\label{subsec:symmetries_KJG}

It can be verified that the Hamiltonian in Eq. (\ref{eq:Ham_rot}) is invariant under the following symmetry transformations,
\begin{eqnarray}
1.&T &:  (S_i^x,S_i^y,S_i^z)\rightarrow (-S_{i}^x,-S_{i}^y,-S_{i}^z)\nn\\
2.& R_aT_a&:  (S_i^x,S_i^y,S_i^z)\rightarrow (S_{i+1}^z,S_{i+1}^x,S_{i+1}^y)\nn\\
3.&R_I I_2&: (S_i^x,S_i^y,S_i^z)\rightarrow (-S_{4-i}^z,-S_{4-i}^y,-S_{4-i}^x),
\label{eq:sym_Jneq0}
\end{eqnarray}
in which $I_2$ is the spatial inversion with inversion center at site $2$  in Fig. \ref{fig:bond_pattern} (b);
and $R_a=R(\hat{z}^{\prime\prime},-2\pi/3)$, $R_I=R(\hat{y}^{\prime\prime},\pi)$ where 
$\hat{x}^{\prime\prime}$, $\hat{y}^{\prime\prime}$ and $\hat{z}^{\prime\prime}$ are related to $\hat{x}^{\prime}$, $\hat{y}^{\prime}$ and $\hat{z}^{\prime}$ via
\bea
(\hat{x}^{\prime\prime}~\hat{y}^{\prime\prime}~\hat{z}^{\prime\prime})=(\hat{x}^\prime~\hat{y}^\prime~\hat{z}^\prime) O,
\label{eq:na_nI}
\eea
in which the orthogonal matrix $O$ is defined as
\bea
O=\left(\begin{array}{ccc}
\frac{1}{\sqrt{6}} & \frac{1}{\sqrt{2}} & \frac{1}{\sqrt{3}}\\
-\sqrt{\frac{2}{3}} & 0 & \frac{1}{\sqrt{3}}\\
\frac{1}{\sqrt{6}} & -\frac{1}{\sqrt{2}} & \frac{1}{\sqrt{3}}
\end{array}
\right).
\label{eq:O_rotate}
\eea
We will call the coordinate frame after the $O$ transformation superimposed on the $U_6$ transformation  as the $OU_6$ frame. 
Throughout this work, $\vec{S}_i$, $\vec{S}_i^\prime$, and $\vec{S}_i^{\prime\prime}$ will be used to denote the spin operators in the original, $U_6$, and $OU_6$ frames, respectively. 

The symmetry group $G^\prime$ of the Hamiltonian $H^\prime$ in Eq. (\ref{eq:sym_Jneq0}) is generated by the symmetry operations in  Eq. (\ref{eq:sym_Jneq0}) as
\begin{eqnarray}
G'=\left< T,R_aT_a,R_II_2\right>.
\label{eq:G_prime}
\end{eqnarray}
Notice that the symmetry groups $G$ and $G^\prime$ must be related by the six-sublattice rotation $U_6$.
Indeed, the generators of $G$ and $G^\prime$ are related by 
\begin{eqnarray}
U_6^{-1} R_aT_a U_6&=&R(\hat{n}_N,\pi)T_a\nn\\ 
U_6^{-1} R_I I_2 U_6&=&T_{2a}^{-1}\cdot R(\hat{n}_N,\pi)T_a\cdot T_aI_0.
\label{eq:U6_G1}
\end{eqnarray}
 
%---------------------------------------------------------------------------------------------------------------------------- 
\section{Couplings induced by electric fields}
\label{sec:coupling_E_field}

In this section, we give the expressions of the spin interactions induced by electric fields,
and discuss the transformation properties of the electric fields under symmetry operations.

\subsection{Dzyaloshinskii-Moriya interaction induced by electric fields}
\label{subsubsec:DM_E_field}

In Kitaev materials, Dzyaloshinskii-Moriya interactions can be induced by applying electric fields.
Consider bond $\gamma$ connecting nearest neighboring sites $i,j$, and the right-handed coordinate system $(\gamma,\alpha,\beta)$.
An electric field is called in-plane if its direction is within the $\alpha\beta$-plane, and out-of-plane if it is parallel with the $\gamma$-axis. 
As shown in Ref. \onlinecite{Furuya2024}, 
an in-plane electric field induces a Dzyaloshinskii-Moriya interaction on bond $\gamma$ in Fig. \ref{fig:bond_pattern} (a) of the form
\bea
D_M^{\text{(in)}}\hat{\gamma}\cdot( \vec{S}_i\times\vec{S}_{j}),
\label{eq:D_Fz}
\eea
in which
\begin{flalign}
&D_M^{\text{(in)}}=\lambda_{D+}^{\text{(in)}}(E_\alpha+E_\beta)+\lambda_{D-}^{\text{(in)}}(E_\alpha-E_\beta),
\label{eq:DMi}
\end{flalign}
where $E_\alpha$ and $E_\beta$ are the electric  field components along the $\alpha$- and $\beta$-directions.
On the other hand, an out-of-plane electric field leads to the following interaction on the $\gamma$ bond \cite{Furuya2024}
\bea
&D_M^{\text{(out)}}(\hat{\alpha}+\hat{\beta})\cdot( \vec{S}_i\times\vec{S}_{j})+\nn\\
&\Gamma^{\prime\text{(out)}} [S_{i}^\gamma (S_{j}^\alpha+S_{j}^\beta)+(S_{i}^\alpha+S_{i}^\beta) S_{j}^\gamma],
\label{eq:D_F_xy}
\eea
in which 
\bea
D_M^{\text{(out)}}&=&\lambda^{\text{(out)}}_D E_\gamma,\notag\\
\Gamma^{\prime\text{(out)}}&=&\lambda^{\text{(out)}}_\Gamma E_\gamma.
\label{eq:DMo}
\eea
In Eq. (\ref{eq:DMi},\ref{eq:DMo}), $\lambda_{D+}^{\text{(in)}},\lambda_{D-}^{\text{(in)}},\lambda^{\text{(out)}}_D,\lambda^{\text{(out)}}_\Gamma$ are related to microscopic parameters including on-site Hubbard interactions, hopping integrals, energy differences between different orbitals, and Hund's coupling, as derived from Ref. \onlinecite{Furuya2024}.
Detailed expressions are included in Appendix \ref{app:expression_constants}.

An estimation of order of magnitudes of the coupling constants in iridates is given in Appendix \ref{app:estimations_E_couplings}.
In accordance with the discussions in Appendix \ref{app:estimations_E_couplings}, 
by setting the unit for electric fields as eV/nm, 
a typical set of values for $\lambda_{D+}^{\text{(in)}},\lambda_{D-}^{\text{(in)}},\lambda_{D}^{\text{(out)}},\lambda_\Gamma^{\text{(out)}}$ can be taken as
\bea
\lambda_{D+}^{\text{(in)}}&=&-0.1\nn\\
\lambda_{D-}^{\text{(in)}}&=&-0.05\nn\\
\lambda_{D}^{\text{(out)}}&=&0.04\nn\\
\lambda_\Gamma^{\text{(out)}}&=&0.05,
\label{eq:numerics_coefficients}
\eea
which will be used in numerical calculations in this work. 
Again we note that Eq. (\ref{eq:numerics_coefficients}) does not give the precise values of the coupling constants in real materials,
but give the correct order of magnitudes.

\subsection{Couplings in six-sublattice-rotated frame}

In the six-sublattice rotated  frame, 
the couplings induced by electric fields are six-site periodic, different from $H^\prime$ in Eq. (\ref{eq:Ham_rot}) having a three-site periodicity. 
In what follows, we use $\tilde{\gamma}\in\{x,y,z,\bar{x},\bar{y},\bar{z}\}$ to denote the bond for couplings involving electric fields,
where the bond pattern for $\tilde{\gamma}$ is shown in Fig. \ref{fig:bond_pattern} (c). 
Similar as before, $\tilde{\alpha},\tilde{\beta}$ are defined as the two spin directions complementary to the direction associated with $\tilde{\gamma}$,
and $(\tilde{\gamma},\tilde{\alpha},\tilde{\beta})$ form a right-handed coordinate system.

As can be verified, after performing the six-sublattice rotation, 
an in-plane electric field in Eq. (\ref{eq:D_Fz}) induces a coupling on bond $\tilde{\gamma}$ in Fig. \ref{fig:bond_pattern} (c) as
\bea
(-)^{i-1}D_M^{\text{(in)}}(-S_i^{\prime\tilde{\alpha}} S_j^{\prime\tilde{\alpha}} + S_i^{\prime\tilde{\beta}} S_j^{\prime\tilde{\beta}}),
\label{eq:Ham_DMi_U6}
\eea
whereas an out-of-plane electric field in Eq. (\ref{eq:D_F_xy}) induces a coupling on bond $\tilde{\gamma}$ as
\begin{flalign}
& (-)^i D_M^{\text{(out)}} [(S_i^{\prime\tilde{\gamma}} S_j^{\prime\tilde{\alpha}}-S_i^{\prime\tilde{\alpha}} S_j^{\prime\tilde{\gamma}})+(S_i^{\prime\tilde{\gamma}} S_j^{\prime\tilde{\beta}}-S_i^{\prime\tilde{\beta}} S_j^{\prime\tilde{\gamma}})]\notag\\
&-\Gamma^{\prime\text{(out)}} [(S_i^{\prime\tilde{\gamma}} S_j^{\prime\tilde{\alpha}}-S_i^{\prime\tilde{\alpha}} S_j^{\prime\tilde{\gamma}})-(S_i^{\prime\tilde{\gamma}} S_j^{\prime\tilde{\beta}}-S_i^{\prime\tilde{\beta}} S_j^{\prime\tilde{\gamma}})].
\label{eq:Ham_DMi_out_U6}
\end{flalign}

\subsection{Symmetry transformations of electric fields} 
\label{ssubsec:sym_transform_uniform_E}

To facilitate symmetry analysis in later sections, we discuss in details how electric field terms transform under symmetry operations. 
We will be interested in symmetry transformation properties in the six-sublattice rotated frame.

We introduce the following terms in the six-sublattice rotated frame,
\begin{flalign}
&\mathcal{H}_{D,\text{odd/even}}^{\prime\text{(in)}}=\sum_{\substack{\langle ij \rangle \in\tilde{\gamma}\,\text{bond} \\ i\in \text{odd/even}}}
(-)^{i-1}(-S_i^{\prime\tilde{\alpha}} S_j^{\prime\tilde{\alpha}} + S_i^{\prime\tilde{\beta}} S_j^{\prime\tilde{\beta}})\nn\\
&\mathcal{H}_{D,\text{odd/even}}^{\prime\text{(out)}}=\nn\\
&\sum_{\substack{\langle ij \rangle \in\tilde{\gamma}\,\text{bond} \\ i\in \text{odd/even}}}
 (-)^i[(S_i^{\prime\tilde{\gamma}} S_j^{\prime\tilde{\alpha}}-S_i^{\prime\tilde{\alpha}} S_j^{\prime\tilde{\gamma}})+(S_i^{\prime\tilde{\gamma}} S_j^{\prime\tilde{\beta}}-S_i^{\prime\tilde{\beta}} S_j^{\prime\tilde{\gamma}})]\nn\\
&\mathcal{H}_{\Gamma,\text{odd/even}}^{\prime\text{(out)}}=\nn\\
&\sum_{\substack{\langle ij \rangle \in\tilde{\gamma}\,\text{bond} \\ i\in \text{odd/even}}}
[-(S_i^{\prime\tilde{\gamma}} S_j^{\prime\tilde{\alpha}}-S_i^{\prime\tilde{\alpha}} S_j^{\prime\tilde{\gamma}})+(S_i^{\prime\tilde{\gamma}} S_j^{\prime\tilde{\beta}}-S_i^{\prime\tilde{\beta}} S_j^{\prime\tilde{\gamma}})]. 
\label{eq:E_uniform_rot2}
\end{flalign}
It is enough to consider how they transform under generators of the symmetry group $G^\prime$ in Eq. (\ref{eq:G_prime}). 
The couplings in Eq. (\ref{eq:E_uniform_rot2}) are apparently invariant under time reversal operation.
Their  transformation properties under $R_aT_a$ and $R_II_2$ can be derived as (for details, see Appendix \ref{app:sym_transform_E_field})
\bea
R_aT_a&:&\mathcal{H}_{D,\text{odd}}^{\prime\text{(in)}}\longleftrightarrow -\mathcal{H}_{D,\text{even}}^{\prime\text{(in)}}\nn\\
&&\mathcal{H}_{D,\text{odd}}^{\prime\text{(out)}}\longleftrightarrow -\mathcal{H}_{D,\text{even}}^{\prime\text{(out)}}\nn\\
&&\mathcal{H}_{\Gamma,\text{odd}}^{\prime\text{(out)}}\longleftrightarrow \mathcal{H}_{\Gamma,\text{even}}^{\prime\text{(out)}},
\label{eq:E_transform_rot_RTa}
\eea
and
\bea
R_II_2&:&\mathcal{H}_{D,\text{odd}}^{\prime\text{(in)}}\longleftrightarrow\mathcal{H}_{D,\text{even}}^{\prime\text{(in)}}\nn\\
&&\mathcal{H}_{D,\text{odd}}^{\prime\text{(out)}}\longleftrightarrow \mathcal{H}_{D,\text{even}}^{\prime\text{(out)}}\nn\\
&&\mathcal{H}_{\Gamma,\text{odd}}^{\prime\text{(out)}}\longleftrightarrow \mathcal{H}_{\Gamma,\text{even}}^{\prime\text{(out)}}.
\label{eq:E_transform_rot_RI}
\eea

%%%%%%%%%%%%%%%%%%%%%%%%%%%%%%%%%%%%%%%%%%%%%%%%%%%%%
\section{Low energy theory without external fields}
\label{sec:Low_energy_theory_zero_field}

In this section, we briefly review the low energy field theory of the model in Eq. (\ref{eq:Ham_1D}) in the absence of electric fields,
which provides a perturbative starting point for analyzing the effects of nonzero fields in later sections. 
Throughout this work, we will consider the parameter region $(K_1<0,\Gamma_1>0,J_1>0)$,
which is relevant for honeycomb iridates. 
We follow the strategy mentioned at the beginning of Sec. \ref{subsec:six-rotation}, by considering the Kitaev-Gamma model first, then treating the Heisenberg coupling $J_1$ as a perturbation, and finally analyzing  the effects of $\Gamma_1^\prime,J_2,K_2$.

%-------------------------------------------- 
\begin{figure}[h]
\includegraphics[width=8cm]{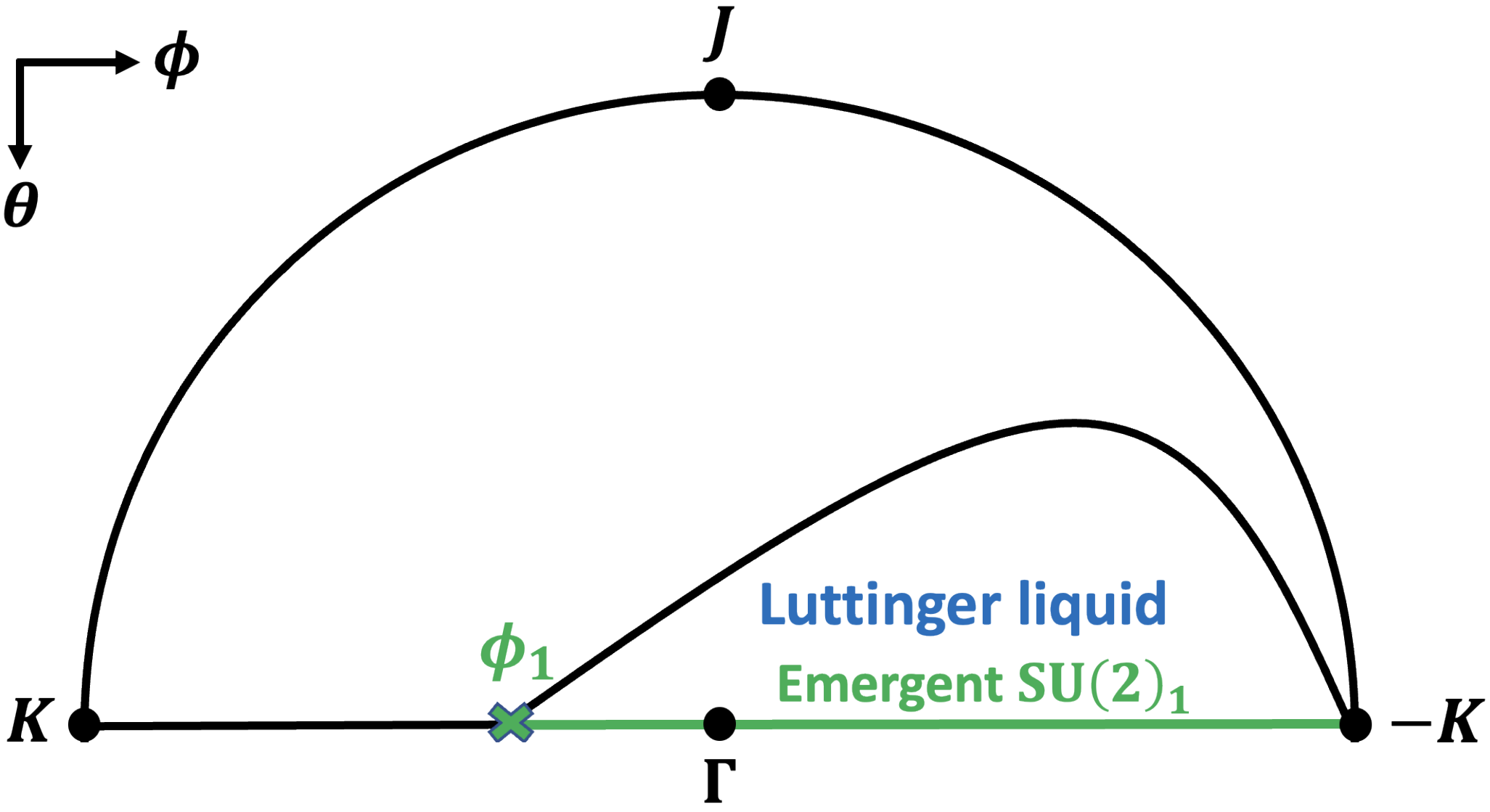}
\caption{
Schematic plot of the Luttinger liquid phase of spin-1/2 $K_1J_1\Gamma_1$ model in the $(K_1<0,\Gamma_1>0,J_1>0)$ region,
in which $K_1,J_1,\Gamma_1$ are parametrized by $\theta,\phi$ in accordance with Eq. (\ref{eq:parametrize_KJG}).
The line marked with green color corresponds to Kitaev-Gamma model, which has an emergent SU(2)$_1$ conformal symmetry at low energies where the transition point $\phi_1$ is $0.33\pi$,   as discussed in Ref. \onlinecite{Yang2020}.
The system remains in the Luttinger liquid phase in the $K_1J_1\Gamma_1\Gamma^\prime_1K_2J_2$ model for small enough $\Gamma_1^\prime,K_2,J_2$.
%The FM$^\prime$ and AFM$^{\prime}$ points have hidden SU(2) symmetries as revealed by the six-sublattice rotation. 
} \label{fig:phase_KHG_1D}
\end{figure}
%--------------------------------------------

\subsection{The Kitaev-Gamma model}

Ref. \onlinecite{Yang2020} shows that in the $(K_1<0,\Gamma_1>0)$ region, the low energy physics of the 1D spin-1/2 Kitaev-Gamma model in the six-sublattice rotated frame is described by the SU(2)$_1$ Wess-Zumino-Witten (WZW) model.
The Hamiltonian of the SU(2)$_1$ WZW model is of the following Sugawara  form 
\bea
H_{\text{WZW}}=\int dx \big[\frac{2\pi}{3}v_0 (\vec{J}^{\prime}_L\cdot \vec{J}^{\prime}_L+\vec{J}'_R\cdot \vec{J^{\prime}_R}),
\label{eq:H_WZW}
\eea
in which $v_0$ is the spin velocity; 
$\vec{J}^{\prime}_L=-\frac{1}{4\pi}\text{tr} [(\partial_z g) g^\dagger \vec{\sigma}]$ and $\vec{J}^{\prime}_R=\frac{1}{4\pi}\text{tr} [g^\dagger (\partial_{\bar{z}} g) \vec{\sigma}]$
are the left and right WZW currents, respectively, 
where the SU(2) matrix $g$ is the WZW primary field, $\sigma^{\alpha}$ ($\alpha=x,y,z$) are the three Pauli matrices,
and $z=\tau+ix$ ($\bar{z}=\tau-ix$) is the holomorphic (anti-holomorphic) coordinate in the imaginary time formalism.
For the Kitaev-Gamma model, 
there exists an additional marginally irrelevant coupling in the low energy field theory, such that the low energy Hamiltonian becomes
\begin{flalign}
H^{\text{Low}}_{K_1\Gamma_1}=\int dx \big[\frac{2\pi}{3}v_0 (\vec{J}^{\prime}_L\cdot \vec{J}^{\prime}_L+\vec{J}'_R\cdot \vec{J^{\prime}_R})-u_0\vec{J}^{\prime}_L\cdot \vec{J}^{\prime}_R\big],
\label{eq:Low_H_KG}
\end{flalign}
where $u_0>0$ is the coupling constant of the marginally irelevant term $\vec{J}^{\prime}_L\cdot \vec{J}^{\prime}_R$.
For later convenience, we define dimerization fluctuation $\epsilon$ and N\'eel order parameter $\vec{N}^{\prime}$ in the $U_6$ frame  as
\bea
\epsilon&=&\text{tr}(g),\nn\\ 
\vec{N}^{\prime}&=&\text{tr}(g\vec{\sigma}).
\eea  

%-------------------------------------------- 
\begin{figure*}[ht]
\begin{center}
\includegraphics[width=14cm]{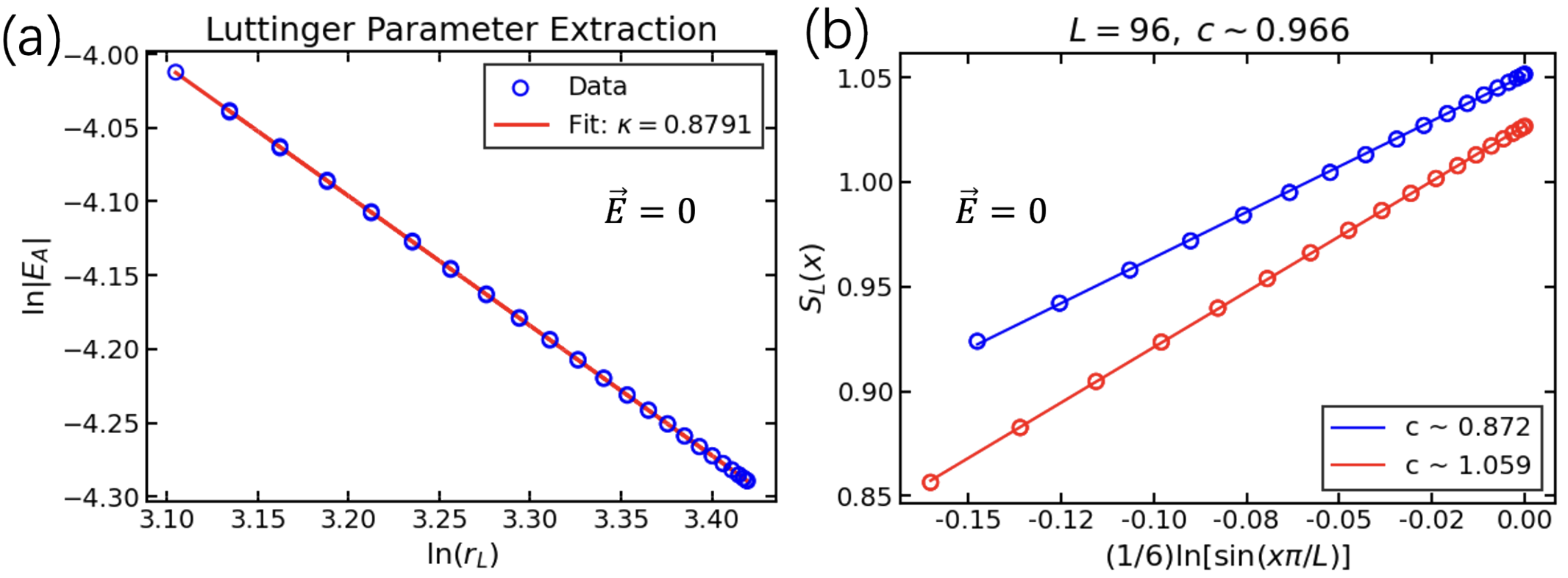}
\caption{(a) Staggered energy density $E_A(r)$ as a function of $r_L$ on a log-log scale and (b) entanglement entropy $S_L(x)$ as a function of $\frac{1}{6}\sin(\frac{\pi x}{L})$ in the absence of electric fields,
where $r_L=\frac{L}{\pi} \sin(\frac{\pi x}{L})$.
In (a,b), DMRG numerics are performed on a system of $L=96$ sites using open boundary conditions,
with bond dimension $m$ and truncation error $\epsilon$ chosen as $m=1000$, $\epsilon=10^{-10}$. 
The parameters of $K_1,J_1,\Gamma_1,\Gamma_1^\prime, K_2,J_2$ in DMRG numerics are taken in accordance with Eq. (\ref{eq:numerics_KJG}) and Eq. (\ref{eq:numerics_coefficients}).
} \label{fig:zero_field}
\end{center}
\end{figure*}
%--------------------------------------------

\subsection{The $K_1J_1\Gamma_1\Gamma^\prime_1K_2J_2$ model}

\subsubsection{The Kitaev-Heisenberg-Gamma model}

In Ref. \onlinecite{Yang2020a}, it has been shown that 
in the $(K_1<0,\Gamma_1>0)$ parameter region with a small $J_1>0$ term, 
the Hamiltonian density of the $K_1J_1\Gamma_1$ model acquires the form
\begin{eqnarray}
H^{\prime\text{Low}}_{K_1J_1\Gamma_1}&=&\int dx\big[\frac{2\pi}{3}v_1 (\vec{J}^{\prime}_L\cdot \vec{J}_L^{\prime}+\vec{J}^{\prime}_R\cdot \vec{J}^{\prime}_R)\nn\\
&&+\nu_1(J_L^{\prime\prime z}-J_R^{\prime\prime z})-u_1\vec{J}_L^{\prime}\cdot \vec{J}_R^{\prime} +u_{1z}J^{\prime z}_LJ^{\prime z}_R\big],\nn\\
\label{eq:low_KJG}
\end{eqnarray}
in which $J_\lambda^{\prime\prime z}=\vec{J}^{\prime}_\lambda\cdot \hat{z}^{\prime\prime}$ ($\lambda=L,R$) is along $z^{\prime\prime}$-direction in the $OU_6$-frame.  
The $\nu_1$ term can be eliminated by a chiral rotation \cite{}, which transforms the low energy Hamiltonian in Eq. (\ref{eq:low_KJG}) to
\begin{eqnarray}
H^{\prime\text{Low}}_{K_1J_1\Gamma_1}&=&\int dx\big[\frac{2\pi}{3}\tilde{v}_1 (\vec{J}^{\prime}_L\cdot \vec{J}_L^{\prime}+\vec{J}^{\prime}_R\cdot \vec{J}^{\prime}_R)\nn\\
&&-\tilde{u}_1\vec{J}_L^{\prime}\cdot \vec{J}_R^{\prime} +\tilde{u}_{1z}J^{\prime z}_LJ^{\prime z}_R\big],
\label{eq:low_KJG_B}
\end{eqnarray}
in which $\tilde{v}_1,\tilde{u}_1,\tilde{u}_{1z}$ are different from $v_1,u_1,u_{1z}$ because of the chiral rotation.
We note that for $\lambda_1>0$, the system has a Luttinger liquid behavior at low energies. 

Alternatively, in the language of Abelian bosonization, the low energy physics can be described by the following Luttinger liquid Hamiltonian 
\bea
H_{LL}=\frac{v}{2} \int dx [\frac{1}{\kappa} (\nabla \varphi)^2 +\kappa (\nabla \theta)^2],
\label{eq:LL_liquid}
\eea
in which the $\theta,\varphi$ fields satisfy the commutation relation $[\varphi(x),\theta(x^\prime)]=\frac{i}{2}\text{sgn}(x^\prime-x)$,
 $v$ is the velocity, and $\kappa$ is the Luttinger liquid parameter.
 Notice that $\kappa$ satisfies  $\kappa=0.5$ for $J=0$ and $0.5<\kappa<1$ in the Luttinger liquid phase for $J> 0$,    as revealed by DMRG numerics in Ref. \onlinecite{Yang2020a}.
We also note that in the Luttinger liquid phase, the system has an emergent U(1) rotation symmetry around the $\hat{z}^{\prime\prime}$-direction,
which is expected from the symmetry operation $R_aT_a$ in Eq. (\ref{eq:sym_Jneq0}). 
 
\subsubsection{General cases}
 
 Since the Luttinger liquid phase is derived from general symmetry and renormalization group (RG) analysis (as discussed in details in Ref. \onlinecite{Yang2020a}),
 the phase is stable against small perturbations.
 In particular, as long as $\Gamma^\prime,K_2,J_2$ are small enough,
 the system remains in the same Luttinger liquid phase. 
 
\subsection{DMRG numerics}

The behavior of the energy density can be used as a test for Luttinger liquids. 
In the Luttinger liquid phase, the energy density $E(r)=\left<H_r\right>$ can be separated into a uniform component $E_U(r)$ and a staggered component $E_A(r)$ in the long distance limit $r\gg 1$ if open boundary conditions are used,
i.e.,
\bea
E(r)=E_U(r)+(-)^r E_A(r),
\label{eq:E_UA}
\eea
in which $r$ is the distance measured from one of the two boundaries of the system,
$H_r$ is the sum of the terms in the Hamiltonian starting with site $r$,
$\left<H_r\right>$ is the ground state expectation value of $H_r$, 
and both $E_U(r)$ and $E_A(r)$ are smooth functions of $r$ on a length scale much larger than the lattice constant.  
In the Luttinger liquid phase, $E_A(r)$ is predicted to behave in the long distance limit as  \cite{Laflorencie2006}
\bea
E_A(r)\propto (r_L)^{-\kappa},
\label{eq:LL_EA}
\eea
in which $\kappa$ is the Luttinger parameter, and 
\bea
r_L=\frac{L}{\pi}\sin(\frac{\pi r}{L}).
\eea

The value of central charge can also serve as a test for Luttinger liquid behavior. 
The central charge can be extracted from calculating entanglement entropy \cite{Calabrese2009}.
The entanglement entropy $S_L(x)$ for a subregion of length $x$ in a finite chain of length $L$ is predicted by conformal field theory (CFT)  as \cite{Calabrese2009}
\begin{align}
    S_L(x) = \lambda\frac{c}{3} \ln \left[ \frac{L}{\pi} \sin\left( \frac{\pi x}{L} \right) \right] + \cdots,
    \label{eq:SL_formula}
\end{align}
in which $c$ is the central charge, 
$\lambda=1$ (and $1/2$) for periodic (and open) boundary conditions 
and ``$\cdots$" represents sub-leading terms in the long distance limit. 
In the Luttinger liquid phase, the central charge is predicted to be $c=1$.

We have numerically calculated the energy density $E(r)$ and entanglement entropy $S_L(x)$ at zero electric field on a system of $L=96$ sites  with open boundary conditions. 
The parameters for $K_1,J_1,\Gamma_1,\Gamma_1^\prime, K_2,J_2$ in DMRG simulations are chosen in accordance with Eq. (\ref{eq:numerics_KJG}). 
Fig. \ref{fig:zero_field} (a) shows $E_A(r)$ as a function of $r_L$ on a log-log scale,
which exhibits a nearly perfect linear relation.
The Luttinger parameter $\kappa$ can be extracted from Fig. \ref{fig:zero_field} (a) using Eq. (\ref{eq:LL_EA}),
which gives $\kappa=0.8791$.
Fig. \ref{fig:zero_field} (b) shows $S_L(x)$ as a function of $\frac{1}{6}\ln [\frac{L}{\pi}\sin(\frac{\pi x}{L})]$ ,
which, apart from an oscillating behavior, also exhibits a linear relation.
The central charge value can be determined as the average of the two slopes in Fig. \ref{fig:zero_field} (b),
which gives $c=0.966$, being very close to $1$.

 %%%%%%%%%%%%%%%%%%%%%%%%%%%%%%%%%%%%%%%%%%%%%%%%%%%%%
\section{Effects of electric fields}
\label{sec:E_fields_effects}

In this section, we investigate the effects of electric fields on Luttinger liquid behaviors of generalized Kitaev spin-1/2 chains. 

An electric field along $x$-direction is out-of-plane (in-plane) on $x$- ($y$-) bonds;
an electric field along $y$-direction is in-plane (out-of-plane) on $x$- ($y$-) bonds;
and an electric field along $z$-direction is in-plane on both $x$- and $y$-bonds,
where the bonds refer to those in Fig. \ref{fig:bond_pattern} (a) in the unrotated frame. 
According to the formulas  in Sec. \ref{subsubsec:DM_E_field}, a uniform electric field $\vec{E}=(E_x,E_y,E_z)$ introduces the following terms in the six-sublattice rotated frame,
\bea
&&r_1(\mathcal{H}^{\prime\text{(in)}}_{D,\text{odd}}+\mathcal{H}^{\prime\text{(in)}}_{D,\text{even}})+r_2(\mathcal{H}^{\prime\text{(in)}}_{D,\text{odd}}-\mathcal{H}^{\prime\text{(in)}}_{D,\text{even}})\nn\\
&+&r_3(\mathcal{H}^{\prime\text{(out)}}_{D,\text{odd}}+\mathcal{H}^{\prime\text{(out)}}_{D,\text{even}})+r_4(\mathcal{H}^{\prime\text{(out)}}_{D,\text{odd}}-\mathcal{H}^{\prime\text{(out)}}_{D,\text{even}})\nn\\
&+&r_5 (\mathcal{H}^{\prime\text{(out)}}_{\Gamma,\text{odd}}+\mathcal{H}^{\prime\text{(out)}}_{\Gamma,\text{even}})+r_6 (\mathcal{H}^{\prime\text{(out)}}_{\Gamma,\text{odd}}-\mathcal{H}^{\prime\text{(out)}}_{\Gamma,\text{even}}),
\label{eq:Ham_rotated_Exyz}
\eea
in which
\bea
r_1&=&\frac{1}{2}[\lambda_{D+}^{\text{(in)}} (E_x+E_y+2E_z)-\lambda_{D-}^{\text{(in)}} (E_x-E_y)]\nn\\
r_2&=&\frac{1}{2}[-\lambda_{D+}^{\text{(in)}} (E_x-E_y)+\lambda_{D-}^{\text{(in)}} (E_x+E_y-2E_z)]\nn\\
r_3&=&\frac{1}{2}\lambda_D^{\text{(out)}}(E_x+E_y)\nn\\
r_4&=&\frac{1}{2}\lambda_D^{\text{(out)}}(E_x-E_y)\nn\\
r_5&=&\frac{1}{2}\lambda_\Gamma^{\text{(out)}}(E_x+E_y)\nn\\
r_6&=&\frac{1}{2}\lambda_\Gamma^{\text{(out)}}(E_x-E_y).
\label{eq:expressions_r1234}
\eea

%-------------------------------------------- 
\begin{figure*}[ht]
\begin{center}
\includegraphics[width=14cm]{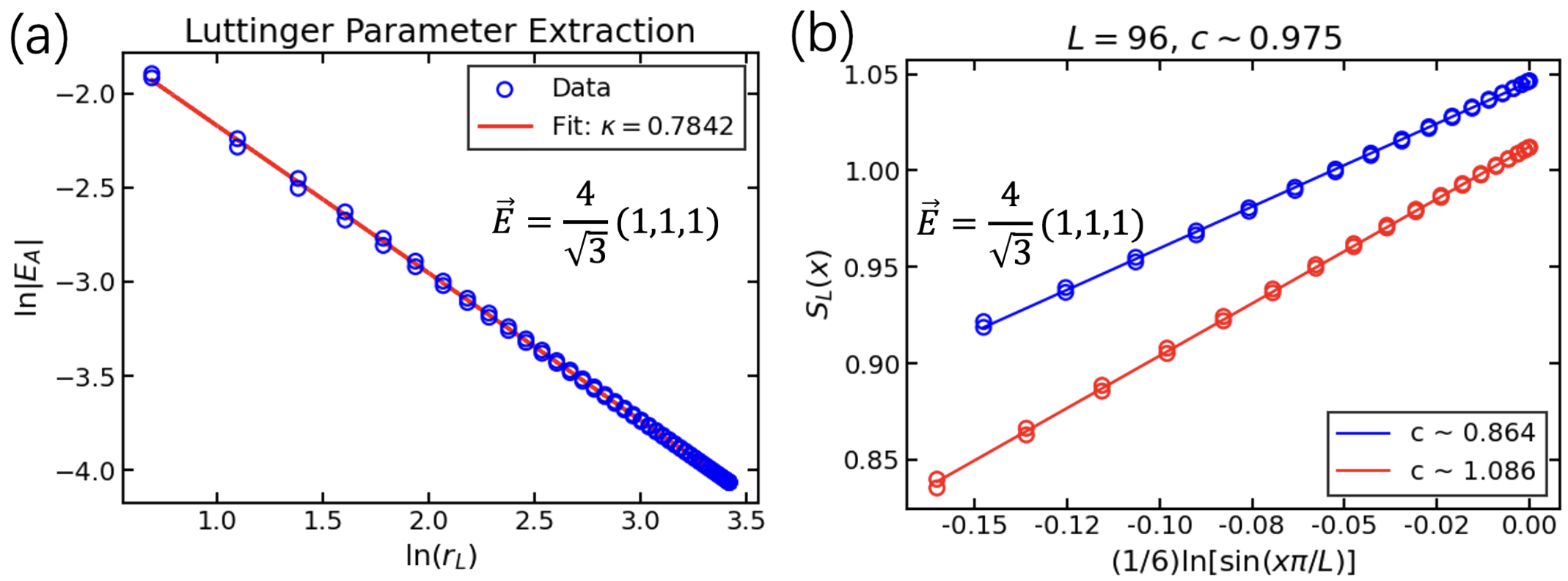}
\caption{(a) Staggered energy density $E_A(r)$ as a function of $r_L$ on a log-log scale and (b) entanglement entropy $S_L(x)$ as a function of $\frac{1}{6}\sin(\frac{\pi x}{L})$ with electric field $\vec{E}=\frac{4}{\sqrt{4}}(1,1,1)$,
where $r_L=\frac{L}{\pi} \sin(\frac{\pi x}{L})$.
In (a,b), DMRG numerics are performed on a system of $L=96$ sites using open boundary conditions,
with bond dimension $m$ and truncation error $\epsilon$ chosen as $m=1000$, $\epsilon=10^{-10}$. 
The parameters of $K_1,J_1,\Gamma_1,\Gamma_1^\prime, K_2,J_2$ in DMRG numerics are taken in accordance with Eq. (\ref{eq:numerics_KJG}) and Eq. (\ref{eq:numerics_coefficients}).
} \label{fig:111_field}
\end{center}
\end{figure*}
%--------------------------------------------

%----------------------------------------------------------------------------------------------------------------------------------------
\subsection{Electric field along $(1,1,1)$-direction}

We start with an electric field along $(1,1,1)$-direction, namely
\bea
\vec{E}_{111}=\frac{1}{\sqrt{3}}E_{111}(\hat{x}+\hat{y}+\hat{z}).
\eea
Using Eq. (\ref{eq:Ham_rotated_Exyz}), the Hamiltonian $H^{\prime}_{E,111}$ in the $U_6$ frame can be obtained as 
\begin{flalign}
&H^{\prime}_{E,111}=H^\prime+\frac{1}{\sqrt{3}}E_{111}\big[
2\lambda_{D+}^{\text{(in)}}(\mathcal{H}^{\prime\text{(in)}}_{D,\text{odd}}+\mathcal{H}^{\prime\text{(in)}}_{D,\text{even}})\nn\\
&+\lambda_{D}^{\text{(out)}}(\mathcal{H}^{\prime\text{(out)}}_{D,\text{odd}}+\mathcal{H}^{\prime\text{(out)}}_{D,\text{even}})+\lambda_{\Gamma}^{\text{(out)}}(\mathcal{H}^{\prime\text{(out)}}_{\Gamma,\text{odd}}+\mathcal{H}^{\prime\text{(out)}}_{\Gamma,\text{even}})
\big],
\label{eq:E111_Ham_Efield}
\end{flalign}
where $H^\prime$ is given in Eq. (\ref{eq:Ham_rot}). 

%----------------------------------------------------------------------
\subsubsection{Luttinger liquid behavior}

According to the discussions in Sec. \ref{ssubsec:sym_transform_uniform_E},
it is straightforward to see that the symmetry group $G_{E,111}^\prime$ of $H^\prime_{E,111}$ in Eq. (\ref{eq:E111_Ham_Efield}) is
\bea
G_{E,111}^\prime=\left< T, R_a^{-1}T_{2a},R_I I_2\right>.
\label{eq:sym_group_G_E_111}
\eea
Using the symmetry transformation properties in Eqs. (\ref{eq:transformT},\ref{eq:transformTa},\ref{eq:transformI},\ref{eq:transformR}), we perform a symmetry analysis to figure out the low energy field theory for the case of an electric field along $(1,1,1)$-direction. 
The SU(2)$_1$ WZW model in Eq. (\ref{eq:H_WZW}) is taken as the unperturbed system. 
For perturbations, in the sense of renormalization group flow, we will only keep relevant and marginal operators, which have scaling dimensions smaller than and equal to two. 

For a symmetry analysis of the low energy theory, it turns out to be easier to work in the $OU_6$ frame,
by defining $\vec{J}^{\prime\prime}$ and $\vec{N}^{\prime\prime}$ as 
\bea
(J^{\prime\prime x},J^{\prime\prime y},J^{\prime\prime z})&=&(J^{\prime x},J^{\prime y},J^{\prime z})O,\nn\\
(N^{\prime\prime x},N^{\prime\prime y},N^{\prime\prime z})&=&(N^{\prime x},N^{\prime y},N^{\prime z})O,
\eea
in which $O$ is defined in Eq. (\ref{eq:O_rotate}).
All symmetry allowed relevant and marginal operators can be classified as follows. 

1) $\epsilon=\text{tr}(g)$ is forbidden by $R_II_2$.

2) $N^{\prime\prime\alpha}$ ($\alpha=x,y,z$) is forbidden by $T$.

3) For $J^{\prime\prime\alpha}_{L/R}$ ($\alpha=x,y,z$),
time reversal requires the combination $J_L^{\prime\prime\alpha}-J_R^{\prime\prime\alpha}$.
Then $R_a^{-1}T_{2a}$ requires $J_L^{\prime\prime z}-J_R^{\prime\prime z}$,
which is invariant under $R_II_2$.

4) For $J^{\prime\prime\alpha}_L\epsilon,J^{\prime\prime\alpha}_R\epsilon$ ($\alpha=x,y,z$),
time reversal requires the combination $(J_L^{\prime\prime\alpha}-J_R^{\prime\prime\alpha})\epsilon$.
Then $R_a^{-1}T_{2a}$ requires $(J_L^{\prime\prime z}-J_R^{\prime\prime z})\epsilon$.
However, $(J_L^{\prime\prime z}-J_R^{\prime\prime z})\epsilon$ changes sign under $R_II_2$, hence forbidden.

5) Within $J^{\prime\prime\alpha}_LN^{\prime\prime\beta},J^{\prime\prime\alpha}_RN^{\prime\prime\beta}$, 
the terms allowed by $T$ and $R_a^{-1}T_{2a}$ are $(\vec{J}^{\prime\prime}_L+\vec{J}^{\prime\prime}_R)\cdot \vec{N}$, $(J_L^{\prime\prime z}+J_R^{\prime\prime z})N^{\prime\prime z}$, and $(J_L^{\prime\prime x}+J_R^{\prime\prime x}) N^{\prime\prime y}-(J_L^{\prime\prime y}+J_R^{\prime\prime y}) N^{\prime\prime x}$,
in which $(J_L^{\prime\prime x}+J_R^{\prime\prime x}) N^{\prime\prime y}-(J_L^{\prime\prime y}+J_R^{\prime\prime y}) N^{\prime\prime x}$ changes sign under $R_II_2$,
hence forbidden.

6) For $J_L^{\prime\prime\alpha} J_R^{\prime\prime\beta}$, the allowed terms by $T$ and $R_a^{-1}T_a$ are
$\vec{J}^{\prime\prime}_L \cdot\vec{J}^{\prime\prime}_R$, $J_L^{\prime\prime z}J_R^{\prime\prime z}$, and $J^{\prime\prime x}_LJ^{\prime\prime y}_R-J^{\prime\prime y}_LJ^{\prime\prime x}_R$,
in which $J^{\prime\prime x}_LJ^{\prime\prime y}_R-J^{\prime\prime y}_LJ^{\prime\prime x}_R$ changes sign under $R_II_2$,
hence forbidden. 

Hence, the low energy Hamiltonian is given by
\begin{flalign}
&H^{\prime\text{Low}}_{E,111}=\frac{2\pi v_2}{3} \int dx (\vec{J}'_L\cdot \vec{J}'_L+\vec{J}'_R\cdot \vec{J}'_R)\nn\\
&+\nu_2\int dx (J_L^{\prime\prime z}-J_R^{\prime\prime z})\nn\\
&+\lambda_2 \int dx (\vec{J}'_L+\vec{J}'_R)\cdot \vec{N}'
+\lambda_{2z} \int dx (J^{\prime\prime z}_L+J^{\prime\prime z}_R)N^{\prime\prime z}\nn\\
&-u_2\int dx \vec{J}'_L\cdot \vec{J}'_R
+u_{2z}\int dx J^{\prime\prime z}_LJ^{\prime\prime z}_R,
\label{eq:low_theory_Ez}
\end{flalign}
in which $v_2,\nu_2,u_2,u_{2z},\lambda_2,\lambda_{2z}$ are the coupling constants of the corresponding terms.
As discussed in Ref. \onlinecite{Yang2025a}, the $\lambda_2,\lambda_{2z}$ terms can be neglected since they are total derivatives, only contributing to boundary terms.
Furthermore, the $\nu_2$ term can be eliminated by a chiral rotation \cite{Garate2010,Gangadharaiah2008,Schnyder2008}.
Removing the $\lambda_2,\lambda_{2z},\nu_2$ terms, the low energy Hamiltonian becomes
\begin{flalign}
&H^{\prime\text{Low}}_{E,111}=\frac{2\pi \tilde{v}_2}{3} \int dx (\vec{J}'_L\cdot \vec{J}'_L+\vec{J}'_R\cdot \vec{J}'_R)\nn\\
&-\tilde{u}_2\int dx \vec{J}'_L\cdot \vec{J}'_R
+\tilde{u}_{2z}\int dx J^{\prime\prime z}_LJ^{\prime\prime z}_R.
\label{eq:low_theory_Ez_2}
\end{flalign}
The low energy physics of $H^{\text{Low}}_{E,111}$ in Eq. (\ref{eq:low_theory_Ez_2}) is the same as an easy-plane AFM XXZ model when $\tilde{u}_{2z}>0$.
Hence, for small enough $E_{z}$, the system remains in the Luttinger liquid phase.  

%----------------------------------------------------------------------
\subsubsection{Numerical results}

We have calculated the staggered energy density $E_A(r)$ and entanglement entropy $S_L(x)$ for  $\vec{E}_{111}=\frac{4}{\sqrt{3}}(1,1,1)$.
DMRG calculations are performed on a system of $L=96$ sites under open boundary conditions, 
with parameters $K_1,J_1,\Gamma_1,\Gamma_1^\prime, K_2,J_2$ and $\lambda_{D+}^{\text{(in)}},\lambda_{D-}^{\text{(in)}},\lambda^{\text{(out)}}_D,\lambda^{\text{(out)}}_\Gamma$ chosen in accordance with Eq. (\ref{eq:numerics_KJG}) and Eq. (\ref{eq:numerics_coefficients}).
$E_A(r)$ vs. $r_L$ on a log-log scale and $S_L(x)$ vs. $\frac{1}{6}\ln [\frac{L}{\pi}\sin(\frac{\pi x}{L})]$ 
are shown in Fig. \ref{fig:111_field} (a) and Fig. \ref{fig:111_field} (b), respectively. 
As can be seen from Fig. \ref{fig:111_field} (a,b),
the DMRG numerical results for both fields are in good agreements with Luttinger liquid behaviors, giving a Luttinger parameter
$\kappa=0.7842$.

Therefore, we see that electric fields along $(1,1,1)$-direction can be used as a method to tune the Luttinger parameter of the system. 
However, such tuning is not quite effective, at least in iridates. 
Comparing Fig. \ref{fig:zero_field} (a) and Fig. \ref{fig:111_field} (a), the Luttinger parameter changes by $10.8\%$ when $\vec{E}=\frac{4}{\sqrt{3}}(1,1,1)$ is applied, corresponding to $\sim 4\times 10^{9}$V/m according to the estimations in Sec. \ref{subsubsec:DM_E_field}. 
Notice that an  $10^9$V/m electric field is already on the order of field values for intrinsic dielectric breakdown in iridate materials \cite{Comin2012}.

%----------------------------------------------------------------------------------------------------------------------------------------
\subsection{Electric field along $z$-direction}

%-------------------------------------------- 
\begin{figure*}[ht]
\begin{center}
\includegraphics[width=14cm]{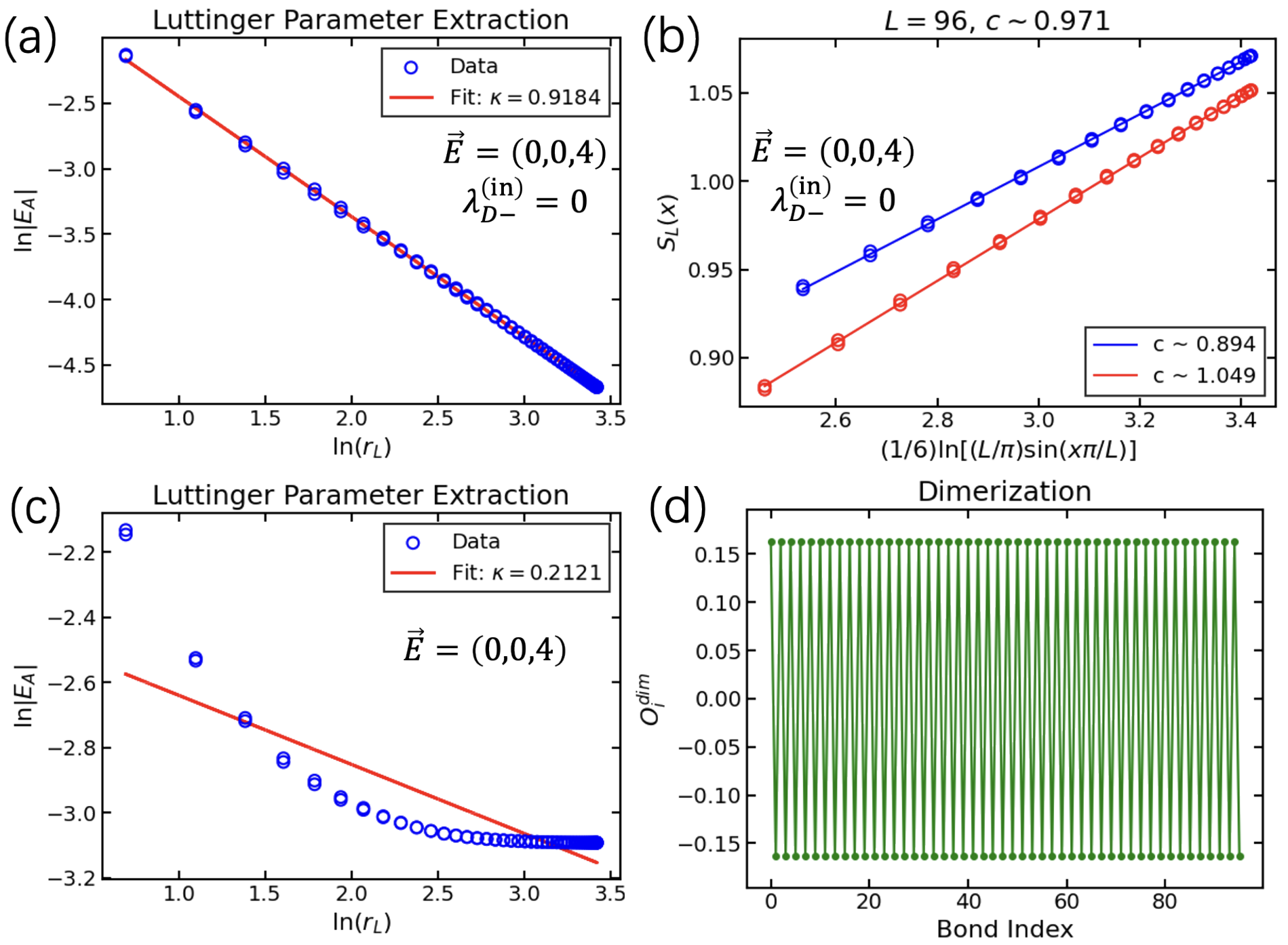}
\caption{(a) Staggered energy density $E_A(r)$ as a function of $r_L$ on a log-log scale and (b) entanglement entropy $S_L(x)$ as a function of $\frac{1}{6}\sin(\frac{\pi x}{L})$ for $\vec{E}=(0,0,4)$ and $\lambda_{D-}^{\text{(0)}}=0$, 
(c) staggered energy density $E_A(r)$ as a function of $r_L$ on a log-log scale and (d) dimerization $O^\prime_{\text{dim},i}$ as a function of site $i$ 
for electric field $\vec{E}=(0,0,4)$ and $\lambda_{D-}^{\text{(0)}}\neq 0$,
where $r_L=\frac{L}{\pi} \sin(\frac{\pi x}{L})$.
In (a,b,c), DMRG numerics are performed on a system of $L=96$ sites using open boundary conditions,
and in (d), DMRG numerics are performed on a system of $L=96$ sites using periodic boundary condition.
Bond dimension $m$ and truncation error $\epsilon$  are chosen as $m=1000$, $\epsilon=10^{-10}$. 
The parameters of $K_1,J_1,\Gamma_1,\Gamma_1^\prime, K_2,J_2$ in DMRG numerics are taken in accordance with Eq. (\ref{eq:numerics_KJG}) and Eq. (\ref{eq:numerics_coefficients}).
} \label{fig:004_field}
\end{center}
\end{figure*}
%--------------------------------------------

Next we consider an electric field along $z$-direction. 
Using Eq. (\ref{eq:Ham_rotated_Exyz}), the Hamiltonian $H^{\prime}_{E,z}$ in the $U_6$ frame can be obtained as 
\bea
H^{\prime}_{E,z}&=&H^\prime+E_z\lambda_{D+}^{\text{(in)}}(\mathcal{H}_{D,\text{odd}}^{\prime\text{(in)}}+\mathcal{H}_{D,\text{even}}^{\prime\text{(in)}})\nn\\
&&-E_z\lambda_{D-}^{\text{(in)}}(\mathcal{H}_{D,\text{odd}}^{\prime\text{(in)}}-\mathcal{H}_{D,\text{even}}^{\prime\text{(in)}}),
\label{eq:Ez_Ham_Efield}
\eea
where $H^\prime$ is given in Eq. (\ref{eq:Ham_rot}). 
Since $J_H\ll U_d,U_p$ in real materials,  $\lambda_{D-}^{\text{(in)}}$ is smaller than $\lambda_{D+}^{\text{(in)}}$ according to Eq. (\ref{eq:DMi_lambda}).
Hence, we will study $\lambda_{D-}^{\text{(in)}}=0$ first, and then treat $\lambda_{D-}^{\text{(in)}}$ as a small perturbation. 

%---------------------------------------------------------------
\subsubsection{Luttinger liquid behavior  for $\lambda_{D-}^{\text{(in)}}=0$}
\label{subsubsec:Low_theory_Ez_lambdam_0}

If $J_H=0$, we have $\lambda_{D-}^{\text{(in)}}=0$, and the Hamiltonian $H^\prime_{E,z0}$ is given by
\bea
H^\prime_{E,z0}=H^\prime+E_z\lambda_{D+}^{\text{(in)}}\big(\mathcal{H}_{D,\text{odd}}^{\prime\text{(in)}}+\mathcal{H}_{D,\text{even}}^{\prime\text{(in)}}\big),
\label{eq:H_Ez_lambdam_0}
\eea
where $H^\prime$ is the Hamiltonian for the spin-1/2 $K_1J_1\Gamma_1\Gamma_1^\prime K_2J_2$ model in the six-sublattice rotated frame given in Eq. (\ref{eq:Ham_rot}). 
According to the discussions in Sec. \ref{ssubsec:sym_transform_uniform_E},
it can be straightforwardly verified that the symmetry group $G_{E,z0}^\prime$ of $H^\prime_{E,z0}$ in Eq. (\ref{eq:H_Ez_lambdam_0}) is
\bea
G_{E,z0}^\prime=\left< T, R_a^{-1}T_{2a},R_I I_2\right>.
\label{eq:Gprime_E_z0}
\eea
Since $G_{E,z0}^\prime$  is the same as $G_{E,111}^\prime$ in Eq. (\ref{eq:sym_group_G_E_111}),
the low energy field theory acquires the same form as Eq. (\ref{eq:low_theory_Ez_2}).
As a result, the system remains in the Luttinger liquid phase. 

%-------------------------------------------- 
\begin{figure*}[ht]
\begin{center}
\includegraphics[width=14cm]{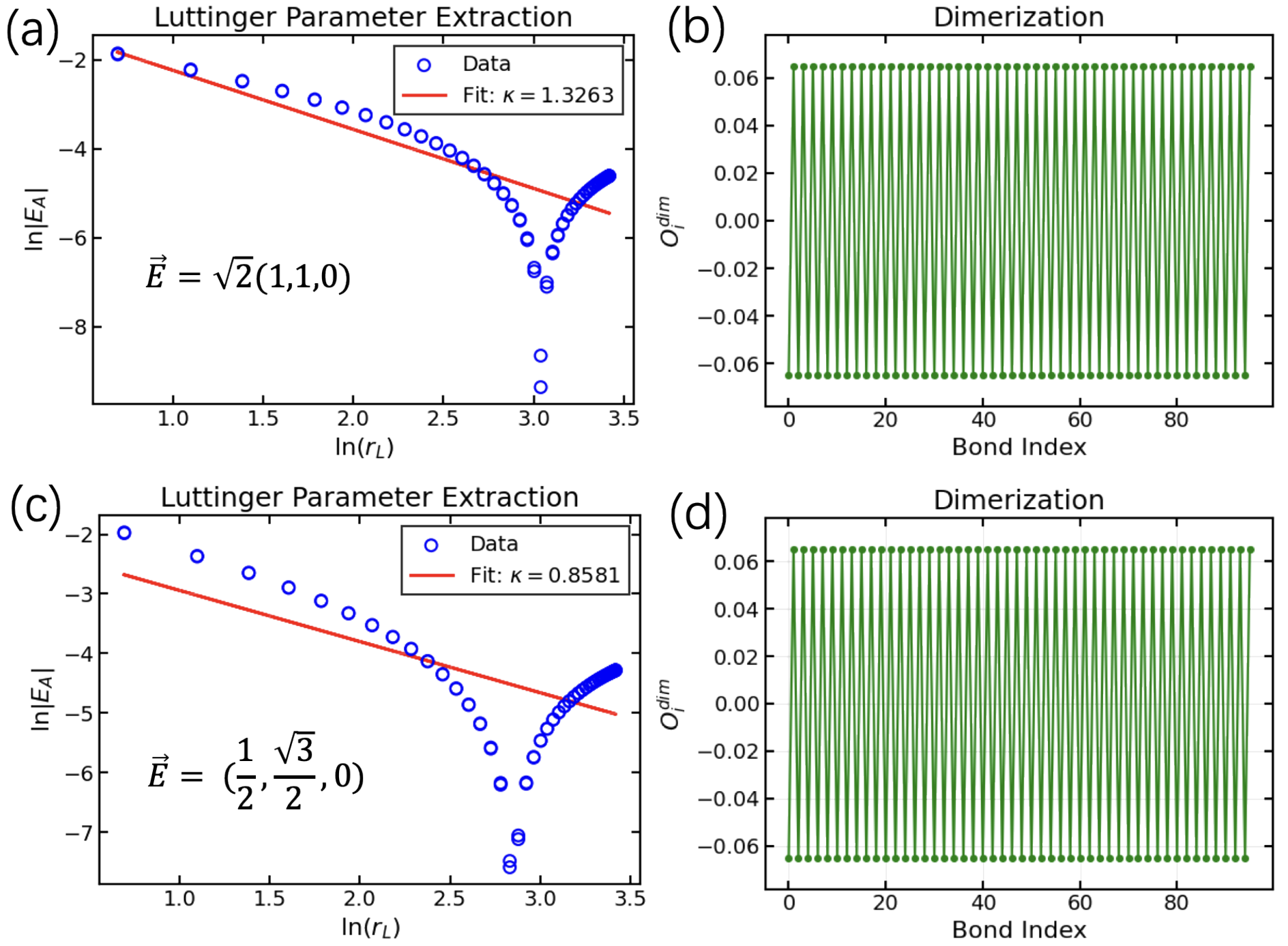}
\caption{(a,c) Staggered energy density $E_A(r)$ as a function of $r_L$ on a log-log scale and (b,d) dimerization $O^\prime_{\text{dim},i}$ as a function of site $i$, with electric field $\vec{E}=\sqrt{2}(1,1,0)$ for (a,b) and $\vec{E}=(\frac{1}{2},\frac{\sqrt{3}}{2},0)$ for (c,d),
where $r_L=\frac{L}{\pi} \sin(\frac{\pi x}{L})$.
In (a,b,c,d), DMRG numerics are performed on a system of $L=96$ sites using open boundary conditions.
Bond dimension $m$ and truncation error $\epsilon$ chosen as $m=1000$, $\epsilon=10^{-10}$. 
The parameters of $K_1,J_1,\Gamma_1,\Gamma_1^\prime, K_2,J_2$ in DMRG numerics are taken in accordance with Eq. (\ref{eq:numerics_KJG}) and Eq. (\ref{eq:numerics_coefficients}).
} \label{fig:220_field}
\end{center}
\end{figure*}
%--------------------------------------------

%---------------------------------------------------------------
\subsubsection{Dimerization for $\lambda_{D-}^{\text{(in)}}\neq 0$}
\label{subsubsec:sym_Ez_lam_minus}

When $J_H\neq 0$, the $\lambda_{D-}^{\text{(in)}}$ term in Eq. (\ref{eq:Ez_Ham_Efield}) is non-vanishing, which  changes sign under $R_I I_2$ according to Eq. (\ref{eq:E_transform_rot_RI}).
Hence, the symmetry group $G'_{E,z}$ for $\lambda_{D-}^{\text{(in)}}\neq 0$ is
\bea
G_{E,z}^\prime=\left< T, R_a^{-1}T_{2a}\right>.
\eea
In this case, compared with the symmetry analysis in Sec. \ref{subsubsec:Low_theory_Ez_lambdam_0}, 
the couplings $\epsilon$, $(J_L^{\prime\prime z}-J_R^{\prime\prime z})\epsilon$, $(J_L^{\prime\prime x}+J_R^{\prime\prime x}) N^{\prime\prime y}-(J_L^{\prime\prime y}+J_R^{\prime\prime y}) N^{\prime\prime x}$, and $J^{\prime\prime x}_LJ^{\prime\prime y}_R-J^{\prime\prime y}_LJ^{\prime\prime x}_R$ are no longer forbidden. 
%The low energy field theory can be written as
%\begin{flalign}
%&H^{\text{Low}}_{E,z}=\frac{2\pi v_2}{3} \int dx (\vec{J}'_L\cdot \vec{J}'_L+\vec{J}'_R\cdot \vec{J}'_R) +a_2\int dx~ \epsilon\nn\\
%&+\nu_2\int dx (J_L^{\prime\prime z}-J_R^{\prime\prime z})+b_2\int dx(J_L^{\prime\prime z}-J_R^{\prime\prime z})\epsilon\nn\\
%&+\lambda_2 \int dx (\vec{J}'_L+\vec{J}'_R)\cdot \vec{N}'
%+\lambda_{2z} \int dx (J^{\prime\prime z}_L+J^{\prime\prime z}_R)N^{\prime\prime z}\nn\\
%&+c_2 \int dx\big[ (J_L^{\prime\prime x}+J_R^{\prime\prime x}) N^{\prime\prime y}-(J_L^{\prime\prime y}+J_R^{\prime\prime y}) N^{\prime\prime x}\big]\nn\\
%&+d_2\int dx \big(J^{\prime\prime x}_LJ^{\prime\prime y}_R-J^{\prime\prime y}_LJ^{\prime\prime x}_R\big)\nn\\
%&-u_2\int dx \vec{J}'_L\cdot \vec{J}'_R
%+u_{2z}\int dx J^{\prime\prime z}_LJ^{\prime\prime z}_R.
%\label{eq:low_theory_Ez_lam_m}
%\end{flalign}

Since $\epsilon$ has the smallest scaling dimension (equal to $1/2$)  among all the symmetry allowed couplings, we only keep $\epsilon$ as the leading relevant perturbations, such that the low energy Hamiltonian $H^{\text{Low}}_{E,z}$  can be simplified as 
\bea
&H^{\prime\text{Low}}_{E,z}= \int dx \big[\frac{2\pi v_3}{3}(\vec{J}'_L\cdot \vec{J}'_L+\vec{J}'_R\cdot \vec{J}'_R)+a_3 \epsilon\big].
\label{eq:H_low_E_z_lam_minus_not_0}
\eea
The effect of the relevant perturbation $\epsilon=\text{tr}(g)$ is to induce a dimerization in the ground states. 
The dimerization order parameter $O^\prime_{\text{dim},i}$ in the six-sublattice rotated frame can be chosen as
\bea
O^\prime_{\text{dim},i}=\vec{S}^\prime_{i-1}\cdot \vec{S}^\prime_{i}-\vec{S}^\prime_{i}\cdot \vec{S}^\prime_{i+1},
\label{eq:Order_dimer}
\eea
and the expectation is that the ground state expectation value of $O^\prime_{\text{dim},i}$ is nonvanishing as long as  $\lambda_{D-}^{\text{(in)}}$  is nonzero.

%----------------------------------------------------------------------
\subsubsection{Numerical results}

For $\lambda_{D-}^{\text{(0)}}=0$, as shown in Fig. \ref{fig:004_field} (a,b), we have calculated the staggered energy density $E_A(r)$ and entanglement entropy $S_L(x)$ with electric field $\vec{E}_{001}=(0,0,4)$  using DMRG calculations on systems of $L=96$ sites under open boundary conditions, 
with parameters $K_1,J_1,\Gamma_1,\Gamma_1^\prime, K_2,J_2$ and $\lambda_{D+}^{\text{(in)}},\lambda^{\text{(out)}}_D,\lambda^{\text{(out)}}_\Gamma$ chosen in accordance with Eq. (\ref{eq:numerics_KJG}) and Eq. (\ref{eq:numerics_coefficients}).
As can be seen from Fig. \ref{fig:004_field} (a,b),
the DMRG numerical results are in good agreement with Luttinger liquid behaviors,
giving a Luttinger parameter $\kappa=0.9184$. 

For $\lambda_{D-}^{\text{(0)}}\neq 0$,
we have calculated the staggered energy density $E_A(r)$ and dimerization $O^\prime_{\text{dim},i}$ for  $\vec{E}_{001}=(0,0,4)$.
$E_A(r)$ vs. $r_L$ on a log-log scale and $O^\prime_{\text{dim},i}$ vs. site $i$
are shown in Fig. \ref{fig:004_field} (c) and Fig. \ref{fig:004_field} (d), respectively. 
DMRG calculations are performed on a system of $L=96$ sites under open boundary condition for $E_A(r)$ and periodic boundary condition for $O^\prime_{\text{dim},i}$, 
with parameters $K_1,J_1,\Gamma_1,\Gamma_1^\prime, K_2,J_2$ and $\lambda_{D+}^{\text{(in)}},\lambda_{D-}^{\text{(in)}},\lambda^{\text{(out)}}_D,\lambda^{\text{(out)}}_\Gamma$ chosen in accordance with Eq. (\ref{eq:numerics_KJG}) and Eq. (\ref{eq:numerics_coefficients}).
As can be seen from Fig. \ref{fig:004_field} (c,d),
no reliable Luttinger parameter can be extracted and the system has a nonzero dimerization.
The DMRG numerical results are in good agreement with a non-Luttinger-liquid behavior. 

%----------------------------------------------------------------------------------------------------------------------------------------
\subsection{Electric field along $(1,1,0)$-direction}

Next, we consider an electric field along $(1,1,0)$-direction, as
\bea
\vec{E}_{110}=\frac{1}{\sqrt{2}}E_{110}(\hat{x}+\hat{y}).
\eea
Using Eq. (\ref{eq:Ham_rotated_Exyz}), the Hamiltonian $H^{\prime}_{E,110}$ in the $U_6$ frame can be obtained as 
\begin{flalign}
&H^{\prime}_{E,110}=H^\prime+\frac{1}{\sqrt{2}}E_{110}\big[\lambda^{\text{(in)}}_{D+} (\mathcal{H}_{D,\text{odd}}^{\prime\text{(in)}}+\mathcal{H}_{D,\text{even}}^{\prime\text{(in)}})\nn\\
&+\lambda^{\text{(in)}}_{D-} (\mathcal{H}_{D,\text{odd}}^{\prime\text{(in)}}-\mathcal{H}_{D,\text{even}}^{\prime\text{(in)}})+\lambda^{\text{(out)}}_{D} (\mathcal{H}_{D,\text{odd}}^{\prime\text{(out)}}+\mathcal{H}_{D,\text{even}}^{\prime\text{(out)}})\nn\\
&+\lambda^{\text{(out)}}_{\Gamma} (\mathcal{H}_{\Gamma,\text{odd}}^{\prime\text{(out)}}+\mathcal{H}_{\Gamma,\text{even}}^{\prime\text{(out)}})\big],
\label{eq:E110_Ham_Efield}
\end{flalign}
where $H^\prime$ is given in Eq. (\ref{eq:Ham_rot}). 

%----------------------------------------------------------------------
\subsubsection{Dimerization}

Since the $\lambda_{D-}^{\text{(in)}}$ term in Eq. (\ref{eq:E110_Ham_Efield}) changes sign under $R_II_2$, the symmetry group $G^\prime_{E,110}$ of $H^{\prime}_{E,110}$ in the $U_6$ frame is 
\bea
G^\prime_{E,110}=\left< T, R_a^{-1}T_{2a}\right>.
\eea
Based on a symmetry analysis, the low energy field theory is the same as Sec. \ref{subsubsec:sym_Ez_lam_minus}, since the symmetry groups are the same. 
In particular, keeping only the most relevant coupling, the low energy Hamiltonian can be written as
\bea
&H^{\prime\text{Low}}_{E,xy}= \int dx \big[\frac{2\pi v_4}{3}(\vec{J}'_L\cdot \vec{J}'_L+\vec{J}'_R\cdot \vec{J}'_R)+a_4 \epsilon\big],
\eea
which has the same form as Eq. (\ref{eq:H_low_E_z_lam_minus_not_0}), but with different values of velocity $v_4$ and coupling constant $a_4$. 
As a result, it is expected  the system is in  a dimerized phase. 

%----------------------------------------------------------------------
\subsubsection{Numerical results}

We have calculated the staggered energy density $E_A(r)$ and dimerization $O^\prime_{\text{dim},i}$ for  $\vec{E}_{110}=\sqrt{2}(1,1,0)$.
$E_A(r)$ vs. on a log-log scale and $O^\prime_{\text{dim},i}$ vs. site $i$
are shown in Fig. \ref{fig:220_field} (a) and Fig. \ref{fig:220_field} (b), respectively. 
DMRG calculations are performed on a system of $L=96$ sites under open boundary condition for $E_A(r)$ and periodic boundary condition for $O^\prime_{\text{dim},i}$, 
with parameters $K_1,J_1,\Gamma_1,\Gamma_1^\prime, K_2,J_2$ and $\lambda_{D+}^{\text{(in)}},\lambda_{D-}^{\text{(in)}},\lambda^{\text{(out)}}_D,\lambda^{\text{(out)}}_\Gamma$ chosen in accordance with Eq. (\ref{eq:numerics_KJG}) and Eq. (\ref{eq:numerics_coefficients}).
As can be seen from Fig. \ref{fig:220_field} (a,b),
no reliable Luttinger parameter can be extracted, and the system has a nonzero dimerization.
The DMRG numerical results are in good agreement with a non-Luttinger-liquid behavior.

%----------------------------------------------------------------------------------------------------------------------------------------
\subsection{General electric fields}

%----------------------------------------------------------------------
\subsubsection{Dimerization }

For electric fields along general directions,
the $r_2$ and $r_4$ terms in Eq. (\ref{eq:Ham_rotated_Exyz}) do not vanish,
since the only case for both $r_2=0$ and $r_4=0$ is $E_x=E_y=E_z$ (namely, an electric field along $(1,1,1)$-direction),
as can be checked from the expressions for $r_i$ ($1\leq i\leq 4$) in Eq. (\ref{eq:expressions_r1234}). 
This means that for general electric fields, the system is not invariant under $R_II_2$,
and as a result, the dimerization operator $\epsilon$ is allowed in the low energy theory.
Hence, the system is dimerized for generic electric fields. 

%----------------------------------------------------------------------
\subsubsection{Numerical results}

We have calculated the staggered energy density $E_A(r)$ and dimerization $O^\prime_{\text{dim},i}$ for  $\vec{E}_{110}=(\frac{1}{2},\frac{\sqrt{3}}{2},0)$.
$E_A(r)$ vs. on a log-log scale and $O^\prime_{\text{dim},i}$ vs. site $i$
are shown in Fig. \ref{fig:220_field} (c) and Fig. \ref{fig:220_field} (d), respectively. 
DMRG calculations are performed on a system of $L=96$ sites under open boundary condition for $E_A(r)$ and periodic boundary condition for $O^\prime_{\text{dim},i}$, 
with parameters $K_1,J_1,\Gamma_1,\Gamma_1^\prime, K_2,J_2$ and $\lambda_{D+}^{\text{(in)}},\lambda_{D-}^{\text{(in)}},\lambda^{\text{(out)}}_D,\lambda^{\text{(out)}}_\Gamma$ chosen in accordance with Eq. (\ref{eq:numerics_KJG}) and Eq. (\ref{eq:numerics_coefficients}).
As can be seen from Fig. \ref{fig:220_field} (c,d),
no reliable Luttinger parameter can be extracted, and the system has a nonzero dimerization.
The DMRG numerical results are in good agreement with a non-Luttinger-liquid behavior.

 %%%%%%%%%%%%%%%%%%%%%%%%%%%%%%%%%%%%%%%%%%%%%%%%%%%%%
\section{Summary} 
\label{sec:summary}
 
In summary, we have studied effects of electric fields in the one-dimensional spin-$1/2$ $K_1J_1\Gamma_1\Gamma_1^\prime K_2J_2$ model in the parameter region $K_1<0,\Gamma_1>0,J>0$,
which is relevant for iridate materials. 
While the system is in a Luttinger liquid phase for zero electric field,
we find that electric fields in general induce nonzero dimerization into the system,
except that the electric field along $(1,1,1)$-direction maintains the Luttinger liquid behavior. 
Our work builds a systematic understanding for electric field effects in generalized Kitaev spin-1/2 chains in the $K_1<0,\Gamma_1>0,J>0$ region,
and provides a starting point for approaching the 2D limit by considering a quasi-1D system of weakly coupled chains.

%%%%%%%%%%%%%%%%%%%%%%%%%%%%%%%%%%%%%%%%%%%
\begin{acknowledgments}

W.Y. and H.W. are supported by the National Natural Science Foundation of
China (Grants No. 12474476)
and the Fundamental Research Funds for the Central Universities.
C.X. acknowledges the supports from MOST grant No. 2022YFA1403900, NSFC No. 12104451, NSFC No. 11920101005, and funds from Strategic Priority Research Program of CAS No. XDB28000000. 
The DMRG calculations in this work were performed using the software package ITensor in  Ref. \onlinecite{ITensor}.

\end{acknowledgments}

%%%%%%%%%%%%%%%%%%%%%%%%%%%%%%%%%%%%%%%%%%%%%%%%%%%%%
%%%%%%%%%%%%%%%%%%%%%%%%%%%%%%%%%%%%%%%%%%%%%%%%%%%%%
\appendix

\begin{widetext}

%%%%%%%%%%%%%%%%%%%%%%%%%%%%%%%%%%%%%%%%%%%%%%%%%%%%%
\section{Electric field induced couplings }
\label{app:expression_constants}

\subsection{Expressions of couplings induced by electric fields }
\label{app:expression_E_couplings}

As derived in details in Ref. \onlinecite{Furuya2024}, $\lambda_{D+}^{\text{(in)}},\lambda_{D-}^{\text{(in)}},\lambda^{\text{(out)}}_D,\lambda^{\text{(out)}}_\Gamma$ in Eq. (\ref{eq:DMi},\ref{eq:DMo}) are given by 
\bea
\lambda_{D+}^{\text{(in)}}&=&-\frac{4IJ_F}{t}\frac{U_d-U_p+\Delta_{dp}}{2(U_d-U_p)+J_H}\nn\\
\lambda_{D-}^{\text{(in)}}&=&-\frac{4IJ_F}{t}\frac{J_H}{2(U_d-U_p)+J_H}\nn\\
\lambda^{\text{(out)}}_D&=&\frac{16 t^3}{3} I \frac{1}{(U_d-U_p+\Delta_{dp})^2}\cdot\frac{J_H}{4(U_d-U_p+\Delta_{dp})^2-J_H^2},\nn\\
\lambda^{\text{(out)}}_\Gamma&=&\frac{32 t^3}{9} I \frac{1}{(U_d-U_p+\Delta_{dp})^2}\cdot \frac{U_d-U_p+\Delta_{dp}}{4(U_d-U_p+\Delta_{dp})^2-J_H^2},
\label{eq:DMi_lambda}
\eea
in which 
$t$ is the hopping integral between $d$- and $p$-orbitals;
$U_d$ and $U_p$ are the strengths of on-site Hubbard interaction on $d$- and $p$-orbitals, respectively;
$J_H$ is the Hund's coupling at $p$-orbitals;
$\Delta_{dp}$ is the orbital energy difference between $d$- and $p$-orbitals;
$I$ is the matrix element of the electric dipole moment between $d$- and $p$-orbitals;
and $J_F$ is given by 
\begin{flalign}
J_F=-\frac{8}{3}t^4 \frac{1}{2(U_d-U_p+\Delta_{dp})^2 [(U_d-U_p+\Delta_{dp})-J_H]}.
\label{eq:expression_JF}
\end{flalign}

\subsection{Estimations of magnitudes of couplings}
\label{app:estimations_E_couplings}

In this appendix, we estimate the magnitudes of various coupling constants.
We will not be very careful about the precise values of the couplings,
but only focusing on their order of magnitudes.  

In iridates, the microscopic parameters in Appendix \ref{app:expression_E_couplings} are given by
\bea
U_d=\text{1.5-2.5eV},~
U_p=\text{4-6eV}, ~
J_H=\text{0.3-0.6eV},~ 
t=\text{0.2-0.3eV}, ~
\Delta_{dp}=\text{2-3eV},~
I=\text{0.5\AA}.
\label{eq:values_U}
\eea
Plugging Eq. (\ref{eq:values_U}) into Eqs. (\ref{eq:DMi_lambda},\ref{eq:expression_JF}), we obtain
\bea
\lambda_{D+}^{\text{(in)}}I&=&-1.51\times 10^{-3} \text{nm} \nn\\
\lambda_{D-}^{\text{(in)}}I&=&-0.76\times 10^{-3} \text{nm} \nn\\
\lambda_{D}^{\text{(out)}}I&=&0.55\times 10^{-3} \text{nm} \nn\\
\lambda_{\Gamma}^{\text{(out)}}I&=&0.74\times 10^{-3} \text{nm}.
\label{eq:numbers_lambda}
\eea
Measuring electric fields in unit of $\text{eV/nm}$, the parameters in Eq. (\ref{eq:numbers_lambda}) can be written as
\bea
\lambda_{D+}^{\text{(in)}}I&\rightarrow&-1.51\text{meV} \nn\\
\lambda_{D-}^{\text{(in)}}I&\rightarrow&-0.76\text{meV}\nn\\
\lambda_{D}^{\text{(out)}}I&\rightarrow&0.55\text{meV}\nn\\
\lambda_{\Gamma}^{\text{(out)}}I&\rightarrow&0.74\text{meV}.
\label{eq:numbers_lambda_2}
\eea

On the other hand, the values of $K_1,J_1,\Gamma_1,\Gamma_1^\prime,K_2,J_2$ are 
\bea
-K_1=\text{10-25meV}, ~
J_1=\text{1-5meV},~
\Gamma_1=\text{5-15meV},~
\Gamma_1^\prime=\text{0.5-3meV},~
K_2=\text{0-1meV},~
J_2=\text{1-5meV}.
\eea
Normalizing $K_1^2+J_1^2+\Gamma_1^2=1$, a typical realistic set of parameters can be chosen as 
\bea
\theta=0.44\pi,~
\phi=0.85\pi,
\eea
and
\bea
\Gamma^\prime=0.1,~
K_2=0.02,~
J_2=0.2,~
\lambda_{D+}^{\text{(in)}}=-0.1,~
\lambda_{D-}^{\text{(in)}}=-0.05,~
\lambda_{D}^{\text{(out)}}=0.04,~
\lambda_\Gamma^{\text{(out)}}=0.05,
\eea
in which $\theta,\phi$ are defined in Eq. (\ref{eq:parametrize_KJG}),
and the unit is on the order of $10$meV.

%%%%%%%%%%%%%%%%%%%%%%%%%%%%%%%%%%%%%%%%%%%%%%%%%%%%%
\section{Explicit forms of the Hamiltonians}
\label{app:explicit_form_E_field}

In this appendix, we include the explicit forms of various terms in the Hamiltonian in both unrotated and six-sublattice rotated frames. 

%------------------------------------------------------------------------------------------------------------------------------------------
\subsection{Hamiltonians in unrotated frame}

Since the Hamiltonian has a two-site periodicity, we only present the terms within the first unit cell which contains two bonds.

%-----------------------------------------------------------
\subsubsection{$K_1J_1\Gamma_1\Gamma^\prime_1K_2J_2$ model}

\bea
1. && K S_1^xS_2^x+J(S_1^xS_2^x+S_1^yS_2^y+S_1^zS_2^z)+\Gamma (S_1^yS_2^z+S_1^zS_2^y)
+\Gamma^\prime (S_1^xS_2^y+S_1^yS_2^x+S_1^xS_2^z+S_1^zS_2^x)\nn\\
&&+K_2 S_1^zS_3^z+J_2(S_1^xS_3^x+S_1^yS_3^y+S_1^zS_3^z)\nn\\
2. && K S_2^yS_3^y+J(S_2^xS_3^x+S_2^yS_3^y+S_2^zS_3^z)+\Gamma (S_2^zS_3^x+S_2^xS_3^z)
+\Gamma^\prime (S_2^yS_3^z+S_2^zS_3^y+S_2^yS_3^x+S_2^xS_3^y)\nn\\
&&+K_2 S_2^zS_4^z+J_2(S_2^xS_4^x+S_2^yS_4^y+S_2^zS_4^z).
\eea

%-----------------------------------------------------------
\subsubsection{Electric fields}

The terms induced by an electric field along $x$-direction are given by
\bea
1. && E_x\big[\lambda_{D}^{\text{(out)}} \left(S_1^{z}S_2^{x}-S_1^{x}S_2^{z}+S_1^{x}S_2^{y}-S_1^{y}S_2^{x}\right)
+\lambda_\Gamma^{\text{(out)}}  \left(S_1^x S_2^y + S_1^y S_2^x + S_1^x S_2^z + S_1^z S_2^x\right)\big]\nn\\
2. &&E_x (\lambda_{D+}^{\text{(in)}}-\lambda_{D-}^{\text{(in)}}) (S_2^{z}S_3^{x}-S_2^{x}S_3^{z}).
\eea

The terms induced by an electric field along $y$-direction are given by
\bea
1. && (\lambda_{D+}^{\text{(in)}}+\lambda_{D-}^{\text{(in)}}) E_y (S_1^{y}S_2^{z}-S_1^{z}S_2^{y})\nn\\
2. &&\lambda_{D}^{\text{(out)}} E_y \left(S_2^{x}S_3^{y}-S_2^{y}S_3^{x}+S_2^{y}S_3^{z}-S_2^{z}S_3^{y}\right)
+\lambda_\Gamma^{\text{(out)}} E_y \left(S_2^y S_3^z + S_2^z S_3^y+ S_2^y S_3^x + S_2^x S_3^y\right). 
\eea

The terms induced by an electric field along $z$-direction are given by
\bea
1. && (\lambda_{D+}^{\text{(in)}}-\lambda_{D-}^{\text{(in)}})E_z (S_1^{y}S_2^{z}-S_1^{z}S_2^{y})\nn\\
2. && (\lambda_{D+}^{\text{(in)}}+\lambda_{D-}^{\text{(in)}})E_z (S_2^{z}S_3^{x}-S_2^{x}S_3^{z}). 
\eea

%------------------------------------------------------------------------------------------------------------------------------------------
\subsection{Hamiltonians in six-sublattice rotated frame}

Since the Hamiltonian has a six-site periodicity, we only present the terms within the first unit cell which contains six bonds.
The $K_1J_1\Gamma_1\Gamma^\prime_1K_2J_2$ model is an exception, which is three-site periodic,
so we only include the first three bonds for this model. 

%-----------------------------------------------------------
\subsubsection{$K_1J_1\Gamma_1\Gamma^\prime_1K_2J_2$ model}

\bea
1.\quad && -KS_1^{\prime x}S_2^{\prime x}
-J\bigl(S_1^{\prime x}S_2^{\prime x}+S_1^{\prime y}S_2^{\prime z}+S_1^{\prime z}S_2^{\prime y}\bigr)
+\Gamma\bigl(S_1^{\prime y}S_2^{\prime y}+S_1^{\prime z}S_2^{\prime z}\bigr)
+\Gamma^\prime\bigl(-S_1^{\prime x}S_2^{\prime z}-S_1^{\prime y}S_2^{\prime x}+S_1^{\prime x}S_2^{\prime y}+S_1^{\prime z}S_2^{\prime x}\bigr)\nn\\
&&
+K_2 S_1^{\prime z}S_3^{\prime x}
+J_2\bigl(S_1^{\prime x}S_3^{\prime y}+S_1^{\prime y}S_3^{\prime z}+S_1^{\prime z}S_3^{\prime x}\bigr)\nn\\[0.6em]
2.\quad && -KS_2^{\prime z}S_3^{\prime z}
-J\bigl(S_2^{\prime x}S_3^{\prime y}+S_2^{\prime z}S_3^{\prime z}+S_2^{\prime y}S_3^{\prime x}\bigr)
+\Gamma\bigl(S_2^{\prime y}S_3^{\prime y}+S_2^{\prime x}S_3^{\prime x}\bigr)
+\Gamma^\prime\bigl(S_2^{\prime z}S_3^{\prime x}+S_2^{\prime y}S_3^{\prime z}-S_2^{\prime z}S_3^{\prime y}-S_2^{\prime x}S_3^{\prime z}\bigr)\nn\\
&&
+K_2S_2^{\prime y}S_4^{\prime z}
+J_2\bigl(S_2^{\prime x}S_4^{\prime y}+S_2^{\prime z}S_4^{\prime x}+S_2^{\prime y}S_4^{\prime z}\bigr)\nn\\[0.6em]
3.\quad && -KS_3^{\prime y}S_4^{\prime y}
-J\bigl(S_3^{\prime y}S_4^{\prime y}+S_3^{\prime z}S_4^{\prime x}+S_3^{\prime x}S_4^{\prime z}\bigr)
+\Gamma\bigl(S_3^{\prime z}S_4^{\prime z}+S_3^{\prime x}S_4^{\prime x}\bigr)
+\Gamma^\prime\bigl(-S_3^{\prime y}S_4^{\prime x}-S_3^{\prime z}S_4^{\prime y}+S_3^{\prime y}S_4^{\prime z}+S_3^{\prime x}S_4^{\prime y}\bigr)\nn\\
&&
+K_2S_3^{\prime x}S_5^{\prime y}
+J_2\bigl(S_3^{\prime y}S_5^{\prime z}+S_3^{\prime z}S_5^{\prime x}+S_3^{\prime x}S_5^{\prime y}\bigr).
\eea

%-----------------------------------------------------------
\subsubsection{Electric fields}

The terms induced by an electric field along $x$-direction are given by
\bea
1.\quad && E_x\Bigl[
\lambda_{D}^{\text{(out)}}\Bigl(S_{1}^{\prime z}S_{2}^{\prime x}-S_{1}^{\prime x}S_{2}^{\prime y}-S_{1}^{\prime x}S_{2}^{\prime z}+S_{1}^{\prime y}S_{2}^{\prime x}\Bigr)
+\lambda_{\Gamma}^{\text{(out)}}\Bigl(-S_{1}^{\prime x}S_{2}^{\prime z}-S_{1}^{\prime y}S_{2}^{\prime x}+S_{1}^{\prime x}S_{2}^{\prime y}+S_{1}^{\prime z}S_{2}^{\prime x}\Bigr)
\Bigr]\nn\\
2.\quad && E_x\Bigl(\lambda_{D+}^{\text{(in)}}-\lambda_{D-}^{\text{(in)}}\Bigr)
\Bigl(S_{2}^{\prime y}S_{3}^{\prime y}-S_{2}^{\prime x}S_{3}^{\prime x}\Bigr)\nn\\
3.\quad && E_x\Bigl[
\lambda_{D}^{\text{(out)}}\Bigl(S_{3}^{\prime x}S_{4}^{\prime y}-S_{3}^{\prime y}S_{4}^{\prime z}-S_{3}^{\prime y}S_{4}^{\prime x}+S_{3}^{\prime z}S_{4}^{\prime y}\Bigr)
+\lambda_{\Gamma}^{\text{(out)}}\Bigl(-S_{3}^{\prime y}S_{4}^{\prime x}-S_{3}^{\prime z}S_{4}^{\prime y}+S_{3}^{\prime y}S_{4}^{\prime z}+S_{3}^{\prime x}S_{4}^{\prime y}\Bigr)
\Bigr]\nn\\
4.\quad && E_x\Bigl(\lambda_{D+}^{\text{(in)}}-\lambda_{D-}^{\text{(in)}}\Bigr)
\Bigl(S_{4}^{\prime z}S_{5}^{\prime z}-S_{4}^{\prime y}S_{5}^{\prime y}\Bigr)\nn\\
5.\quad && E_x\Bigl[
\lambda_{D}^{\text{(out)}}\Bigl(S_{5}^{\prime y}S_{6}^{\prime z}-S_{5}^{\prime z}S_{6}^{\prime x}-S_{5}^{\prime z}S_{6}^{\prime y}+S_{5}^{\prime x}S_{6}^{\prime z}\Bigr)
+\lambda_{\Gamma}^{\text{(out)}}\Bigl(-S_{5}^{\prime z}S_{6}^{\prime y}-S_{5}^{\prime x}S_{6}^{\prime z}+S_{5}^{\prime z}S_{6}^{\prime x}+S_{5}^{\prime y}S_{6}^{\prime z}\Bigr)
\Bigr]\nn\\
6.\quad && E_x\Bigl(\lambda_{D+}^{\text{(in)}}-\lambda_{D-}^{\text{(in)}}\Bigr)
\Bigl(S_{6}^{\prime x}S_{7}^{\prime x}-S_{6}^{\prime z}S_{7}^{\prime z}\Bigr)\nn\\
\eea

The terms induced by an electric field along $y$-direction are given by
\bea
1.\quad && E_y(\lambda_{D+}^{\text{(in)}}+\lambda_{D-}^{\text{(in)}}) 
\Bigl(S_{1}^{\prime y}S_{2}^{\prime y}-S_{1}^{\prime z}S_{2}^{\prime z}\Bigr)\nn\\
2.\quad && E_y\Bigl[
\lambda_{D}^{\text{(out)}}\Bigl(-S_{2}^{\prime x}S_{3}^{\prime z}+S_{2}^{\prime z}S_{3}^{\prime y}+S_{2}^{\prime z}S_{3}^{\prime x}-S_{2}^{\prime y}S_{3}^{\prime z}\Bigr)
+\lambda_{\Gamma}^{\text{(out)}}\Bigl(S_{2}^{\prime z}S_{3}^{\prime x}+S_{2}^{\prime y}S_{3}^{\prime z}-S_{2}^{\prime z}S_{3}^{\prime y}-S_{2}^{\prime x}S_{3}^{\prime z}\Bigr)
\Bigr]\nn\\
3.\quad && E_y(\lambda_{D+}^{\text{(in)}}+\lambda_{D-}^{\text{(in)}})
\Bigl(S_{3}^{\prime z}S_{4}^{\prime z}-S_{3}^{\prime x}S_{4}^{\prime x}\Bigr)\nn\\
4.\quad && E_y\Bigl[
\lambda_{D}^{\text{(out)}}\Bigl(-S_{4}^{\prime y}S_{5}^{\prime x}+S_{4}^{\prime x}S_{5}^{\prime z}+S_{4}^{\prime x}S_{5}^{\prime y}-S_{4}^{\prime z}S_{5}^{\prime x}\Bigr)
+\lambda_{\Gamma}^{\text{(out)}}\Bigl(S_{4}^{\prime x}S_{5}^{\prime y}+S_{4}^{\prime z}S_{5}^{\prime x}-S_{4}^{\prime x}S_{5}^{\prime z}-S_{4}^{\prime y}S_{5}^{\prime x}\Bigr)
\Bigr]\nn\\
5.\quad && E_y(\lambda_{D+}^{\text{(in)}}+\lambda_{D-}^{\text{(in)}})
\Bigl(S_{5}^{\prime x}S_{6}^{\prime x}-S_{5}^{\prime y}S_{6}^{\prime y}\Bigr)\nn\\
6.\quad && E_y\Bigl[
\lambda_{D}^{\text{(out)}}\Bigl(-S_{6}^{\prime z}S_{7}^{\prime y}+S_{6}^{\prime y}S_{7}^{\prime x}+S_{6}^{\prime y}S_{7}^{\prime z}-S_{6}^{\prime x}S_{7}^{\prime y}\Bigr)
+\lambda_{\Gamma}^{\text{(out)}}\Bigl(S_{6}^{\prime y}S_{7}^{\prime z}+S_{6}^{\prime x}S_{7}^{\prime y}-S_{6}^{\prime y}S_{7}^{\prime x}-S_{6}^{\prime z}S_{7}^{\prime y}\Bigr)
\Bigr].
\eea

The terms induced by an electric field along $z$-direction are given by
\bea
1.\quad && E_z(\lambda_{D+}^{\text{(in)}}-\lambda_{D-}^{\text{(in)}})
\Bigl(S_{1}^{\prime y}S_{2}^{\prime y}-S_{1}^{\prime z}S_{2}^{\prime z}\Bigr)\nn\\
2.\quad && E_z(\lambda_{D+}^{\text{(in)}}+\lambda_{D-}^{\text{(in)}})
\Bigl(S_{2}^{\prime y}S_{3}^{\prime y}-S_{2}^{\prime x}S_{3}^{\prime x}\Bigr)\nn\\
3.\quad && E_z(\lambda_{D+}^{\text{(in)}}-\lambda_{D-}^{\text{(in)}})
\Bigl(S_{3}^{\prime z}S_{4}^{\prime z}-S_{3}^{\prime x}S_{4}^{\prime x}\Bigr)\nn\\
4.\quad && E_z(\lambda_{D+}^{\text{(in)}}+\lambda_{D-}^{\text{(in)}})
\Bigl(S_{4}^{\prime z}S_{5}^{\prime z}-S_{4}^{\prime y}S_{5}^{\prime y}\Bigr)\nn\\
5.\quad && E_z(\lambda_{D+}^{\text{(in)}}-\lambda_{D-}^{\text{(in)}})
\Bigl(S_{5}^{\prime x}S_{6}^{\prime x}-S_{5}^{\prime y}S_{6}^{\prime y}\Bigr)\nn\\
6.\quad && E_z(\lambda_{D+}^{\text{(in)}}+\lambda_{D-}^{\text{(in)}})
\Bigl(S_{6}^{\prime x}S_{7}^{\prime x}-S_{6}^{\prime z}S_{7}^{\prime z}\Bigr).
\eea

%------------------------------------------------------------------------------------------------------------------------------------------
\section{Symmetry transformation properties of electric field terms}
\label{app:sym_transform_E_field}

It is easier to analyze the transformations in the unrotated frame, since the periodicity of the bond pattern is shorter (i.e., two in the unrotated frame rather than six in the $U_6$ frame). 

According to Eq. (\ref{eq:U6_G1}), $G^\prime$ in Eq. (\ref{eq:G_prime}) can be written as 
\begin{eqnarray}
G^\prime&=&U_6GU^{-1}_6\nn\\
&=&U_6\left< T,R(\hat{n}_N,\pi)T_a,R(\hat{n}_N,\pi)I_0\right>U^{-1}_6.
\label{eq:G_prime_2}
\end{eqnarray}
As a result, for the purpose of knowing how electric field terms transform under $T,R_aT_a,R_II_2$ in the $U_6$ frame,
it is enough to work out how they transform under $T,R(\hat{n}_N,\pi)T_a,R(\hat{n}_N,\pi)I_0$ in the unrotated frame.
Apparently, any uniform electric field is invariant under $T$ and $T_{2a}$.
We will consider how they transform under $R(\hat{n}_N,\pi)I_0$ and $R(\hat{n}_N,\pi)T_a$. 

We introduce the following terms in the unrotated frame,
\bea
\mathcal{H}_{D,\text{odd/even}}^{\text{(in)}}&=&\sum_{\substack{\langle ij \rangle \in\gamma\,\text{bond} \\ i\in \text{odd/even}}}(S_i^\alpha S_{i+1}^\beta-S_i^\beta S_{i+1}^\alpha)\nn\\
\mathcal{H}_{D,\text{odd/even}}^{\text{(out)}}
&=&\sum_{\substack{\langle ij \rangle \in\gamma\,\text{bond} \\ i\in \text{odd/even}}} (S_i^\beta S_{i+1}^\gamma-S_i^\gamma S_{i+1}^\beta
+S_i^\gamma S_{i+1}^\alpha-S_i^\alpha S_{i+1}^\gamma)\nn\\
\mathcal{H}_{\Gamma,\text{odd/even}}^{\text{(out)}}
&=&\sum_{\substack{\langle ij \rangle \in\gamma\,\text{bond} \\ i\in \text{odd/even}}}(S_i^\gamma S_{i+1}^\alpha+S_i^\alpha S_{i+1}^\gamma+S_i^\gamma S_{i+1}^\beta+S_i^\alpha S_{i+1}^\gamma). 
\label{eq:Ham_Efields_DM_in_out}
\eea
It can be verified that the couplings in Eq. (\ref{eq:Ham_Efields_DM_in_out}) transform as 
\bea
R(\hat{n}_N,\pi)T_a&:&\mathcal{H}_{D,\text{odd}}^{\text{(in)}}\longleftrightarrow -\mathcal{H}_{D,\text{even}}^{\text{(in)}}\nn\\
&&\mathcal{H}_{D,\text{odd}}^{\text{(out)}}\longleftrightarrow -\mathcal{H}_{D,\text{even}}^{\text{(out)}}\nn\\
&&\mathcal{H}_{\Gamma,\text{odd}}^{\text{(out)}}\longleftrightarrow \mathcal{H}_{\Gamma,\text{even}}^{\text{(out)}},
\label{eq:E_transform_unrot_RTa}
\eea
and
\bea
R(\hat{n}_N,\pi)I_0&:&\mathcal{H}_{D,\text{odd}}^{\text{(in)}}\longleftrightarrow\mathcal{H}_{D,\text{even}}^{\text{(in)}}\nn\\
&&\mathcal{H}_{D,\text{odd}}^{\text{(out)}}\longleftrightarrow \mathcal{H}_{D,\text{even}}^{\text{(out)}}\nn\\
&&\mathcal{H}_{\Gamma,\text{odd}}^{\text{(out)}}\longleftrightarrow \mathcal{H}_{\Gamma,\text{even}}^{\text{(out)}}.
\label{eq:E_transform_unrot_RI}
\eea

Denote 
\bea
\mathcal{H}_{D,\text{odd/even}}^{\prime\text{(in)}}&=&U_6\mathcal{H}_{D,\text{odd/even}}^{\text{(in)}}U_6^{-1},\nn\\
\mathcal{H}_{D,\text{odd/even}}^{\prime\text{(out)}}&=&U_6\mathcal{H}_{D,\text{odd/even}}^{\text{(out)}}U_6^{-1},\nn\\
\mathcal{H}_{\Gamma,\text{odd/even}}^{\prime\text{(out)}}&=&U_6\mathcal{H}_{\Gamma,\text{odd/even}}^{\text{(out)}}U_6^{-1},
\label{eq:E_uniform_rot}
\eea
which are exactly Eq. (\ref{eq:E_uniform_rot2}).
Explicit forms of the terms in Eq. (\ref{eq:E_uniform_rot}) in the $U_6$ frame are included in Appendix \ref{app:explicit_form_E_field}.
All terms in Eq. (\ref{eq:E_uniform_rot}) are all invariant under $U_6TU^{-1}_6=T$ and $U_6T_{2a}U^{-1}_6=R_a^{-1}T_{2a}$.
Furthermore, their transformation properties under $R_aT_a$ and $R_II_2$ can be obtained from Eqs. (\ref{eq:E_transform_unrot_RTa},\ref{eq:E_transform_unrot_RI}) as Eqs. (\ref{eq:E_transform_rot_RTa},\ref{eq:E_transform_rot_RI}).

%------------------------------------------------------------------------------------------------------------------------------------------
\section{Transformation rules of WZW fields}

The transformation properties of the WZW fields $g$ and $\vec{J}^{\prime}_L,\vec{J}^{\prime}_R$ under time reversal $T$, spatial translation $T_a$, spatial inversion $I$ with inversion center located at sites, and global spin rotation $R\in SO(3)$ are given by
\begin{eqnarray}
T: &\epsilon(x)\rightarrow \epsilon(x), &\vec{N}^{\prime}(x)\rightarrow -\vec{N}^{\prime}(x),\nn\\
&\vec{J}^{\prime}_L(x)\rightarrow -\vec{J}^{\prime}_R(x), &\vec{J}^{\prime}_R(x)\rightarrow -\vec{J}^{\prime}_L(x), 
\label{eq:transformT}
\end{eqnarray}
\begin{eqnarray}
T_a: &\epsilon(x)\rightarrow -\epsilon(x), &\vec{N}^{\prime}(x)\rightarrow -\vec{N}^{\prime}(x),\nn\\
&\vec{J}^{\prime}_L(x)\rightarrow \vec{J}^{\prime}_L(x), &\vec{J}^{\prime}_R(x)\rightarrow \vec{J}^{\prime}_R(x), 
\label{eq:transformTa}
\end{eqnarray}
\begin{eqnarray}
I: & \epsilon(x)\rightarrow -\epsilon(-x), &\vec{N}^{\prime}(x)\rightarrow \vec{N}^{\prime}(-x),\nn\\
&\vec{J}^{\prime}_L(x)\rightarrow \vec{J}^{\prime}_R(-x), &\vec{J}^{\prime}_R(x)\rightarrow \vec{J}^{\prime}_L(-x), 
\label{eq:transformI}
\end{eqnarray}
\begin{eqnarray}
R: &\epsilon(x)\rightarrow \epsilon(x), &N^{\prime\alpha}(x)\rightarrow R^{\alpha}_{\,\,\beta}N^{\prime\beta}(x),\nn\\
&J^{\prime\alpha}_L(x)\rightarrow R^{\alpha}_{\,\,\beta} J^{\prime\beta}_L(x), &J^{\prime\alpha}_R(x)\rightarrow R^{\alpha}_{\,\,\beta}J^{\prime\beta}_R(x).
\label{eq:transformR}
\end{eqnarray}
in which $x$ is the spatial coordinate; $R^{\alpha}_{\,\,\beta}$ ($\alpha,\beta=x,y,z$) is the matrix element of the $3\times 3$ rotation matrix $R$ at position $(\alpha,\beta)$.

\end{widetext}
%%%%%%%%%%%%%%%%%%%%%%%%%%%%%%%%%%%%%%%%%%%%%%%

%%%%%%%%%%%%%%%%%%%%%%%%%%%%%%%%%%%%%%%

\end{document}